**Title: Mammalian gonadal development: origins, specification and differentiation of ovary and testis cell lineages.**

**Authors:** Serge Nef[1,2,*], Maëva Guy[1,2], Marie-Christine Chaboissier[3,*].

**Affiliations:**
[1]Department of Genetic Medicine and Development, University of Geneva, 1211 Geneva, Switzerland.
[2]iGE3, Institute of Genetics and Genomics of Geneva, University of Geneva, Switzerland.
[3]Université Côte d'Azur, Inserm, CNRS, Institut de Biologie Valrose (iBV), 06108 Nice, France.

[*]Corresponding Authors: Email: Serge.Nef@unige.ch, Marie-Christine.Chaboissier@univ-cotedazur.fr

**Abstract:**
Testicular and ovarian development requires the coordinated formation and assembly of multiple cell types that enable these organs to fulfill their roles in sexual differentiation, reproduction, and endocrine function. A defining feature of gonadal development is that both testes and ovaries arise from a common bipotential primordium, the genital ridge, composed of multipotent progenitors. In this Review, we synthesize current knowledge on the origin, specification, differentiation, and function of the major fetal gonadal somatic lineages, drawing primarily on mouse and human studies. We describe how the adrenogonadal primordium (AGP) gives rise to the gonadal primordium (GP), how coelomic epithelium-derived progenitors generate the supporting, steroidogenic/stromal, and supporting-like cell lineages, and how these progenitors become progressively fate-restricted. We highlight recent evidence for early lineage priming within spatially organized domains of the coelomic epithelium. We further examine the molecular pathways and cellular processes governing sex-specific differentiation of these lineages and discuss remaining gaps in our understanding of how bipotential gonads differentiate into testes or ovaries, and their implication for human health.

## Introduction

Sexual reproduction is fundamental to most animal species, enabling the generation of biological diversity and novelty from one generation to the next. This process relies on the production of haploid gametes, sperm and oocytes, and on the existence of two distinct sexes, each with specialized reproductive organs and functions. However, two sexes imposes significant complexity to development, as embryogenesis must generate two physiologically distinct organisms within a single species. In eutherian mammals, sex is determined by chromosomes, with XY individuals developing testes and XX individuals forming ovaries. Beyond gamete production, the gonads function as endocrine organs that orchestrate the development of the reproductive system, ensuring sexual dimorphism and fertility. As such, gonadal development plays a central role in sexual development and reproductive biology.

The process by which the bipotential gonad differentiates into either a testis or an ovary is known as primary or gonadal sex determination. Once formed, the gonads direct phenotypic sex through the secretion of hormones that drive the development of internal and external reproductive structures, culminating in reproductive maturation and gamete production at puberty. Gonadal sex determination represents a central paradigm in developmental biology, as it involves the sexually dimorphic differentiation of a common embryonic structure, the genital ridge. This primordium is considered bipotential, as it has the capacity to develop into either a testis or an ovary. Despite their shared origin and analogous functions, the testes and ovaries diverge profoundly in structure and function, reflecting distinct cellular organizations and gene regulatory programs.

Gonadal differentiation is unusual in several respects. Many cell lineages are initially bipotential and their differentiation is highly interdependent. In particular, the supporting lineage, Sertoli cells in the testis and granulosa cells in the ovary, acts as a central organizer coordinating the sexual differentiation of other cell types including germ cells. Gonadal formation also involves cells of distinct origins: locally derived cells from the coelomic epithelium (CE) and the mesonephros, and primordial germ cells originating from the caudal epiblast. Together, these features require tight coordination between lineage specification, cell differentiation, tissue morphogenesis, and intercellular signaling.

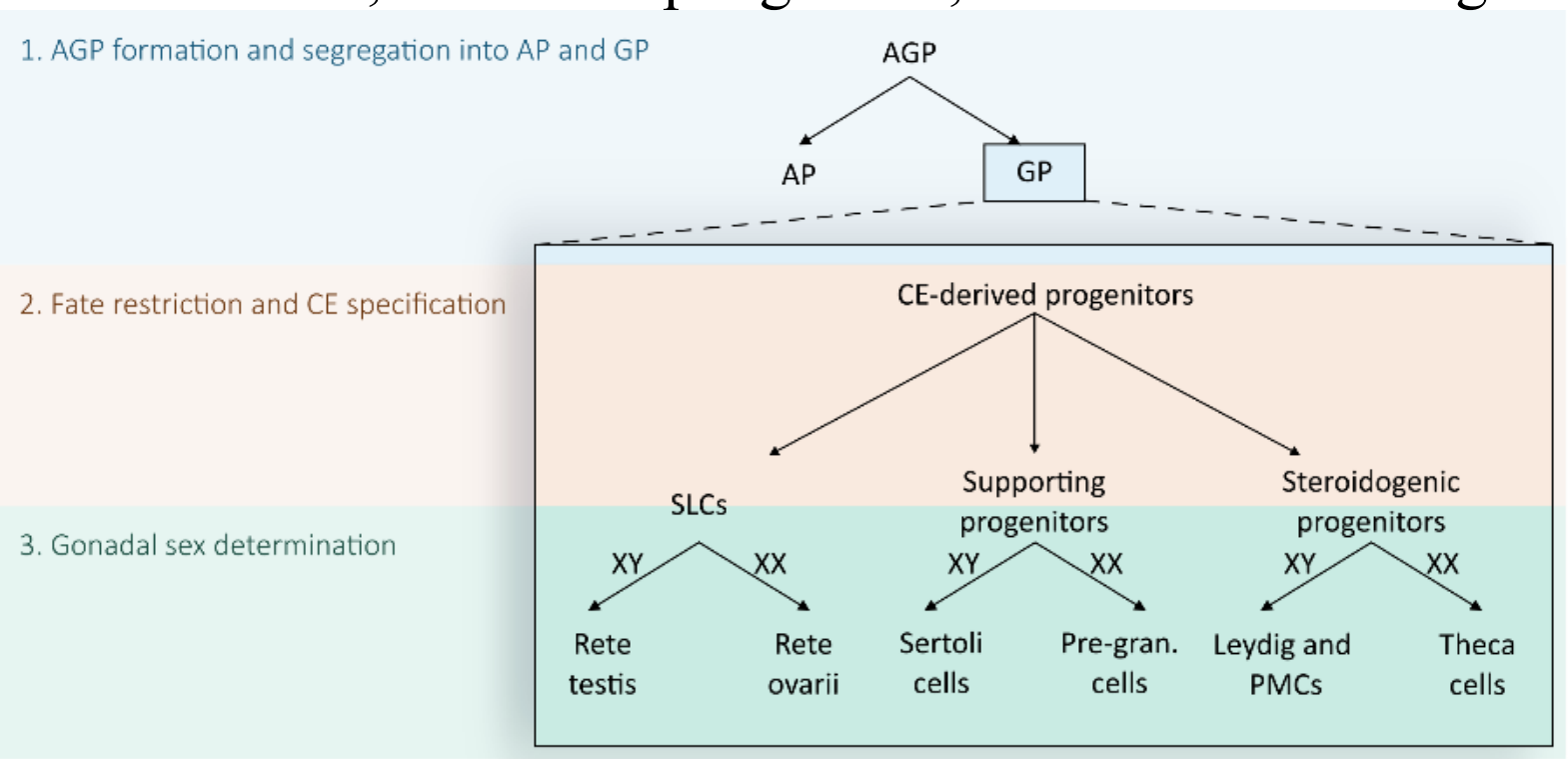

**Figure 1. A three-phase model of gonadal development and sex determination.** Abbreviations: CE, coelomic epithelium; AGP, adreno-gonadal primordium; AP, adrenal primordium; GP, gonadal primordium, SLCs, supporting-like cells; PMCs, peritubular myoid cells.

Gonadal development can be broadly divided into three sequential phases (**Figure 1**). The first stage involves the formation of the adrenogonadal primordium (AGP) and its segregation into adrenal (AP) and gonadal (GP) primordia. The second stage corresponds to the restriction of multipotent CE progenitors and their specification into distinct somatic progenitor populations that give rise to the supporting, steroidogenic, and supporting-like lineages. The third phase is gonadal sex determination, during which these progenitors differentiate in a sex-specific manner. While most studies have focused on this final step, particularly in the context of differences of sex development (DSDs), earlier processes governing progenitor specification and gonad formation have received less attention. These upstream events are likely to be equally critical and may underlie many cases of DSDs, infertility, and other gonadal pathologies.

In this review, we synthesize current knowledge on the origin, specification, differentiation, and function of the main somatic lineages that form the fetal gonad. We present recent findings on how these lineages acquire their identity and how early developmental events shape ovarian and testicular development. We highlight key molecular pathways and genetic regulators underlying gonadal development, primary sex determination, and differentiation, with a particular emphasis on human and mouse studies, including *in vivo* transgenic and knockout models. Finally, we identify remaining gaps and outline future directions to better understand how bipotential gonads differentiate into testes or ovaries and its relevance to human health.

**A common origin for the gonads and adrenal gland: the adreno-gonadal primordium (AGP)**

The adrenal cortex and the gonads share a common developmental origin in the AGP (Bandiera et al., 2013; Val et al., 2007; Zubair et al., 2006; Zubair et al., 2008). While the AGP was historically thought to derive from the intermediate mesoderm (Fraser, 1950), recent single-cell transcriptomic studies support a lateral plate mesoderm origin (Neirijnck et al., 2023; Qiu et al., 2024), consistent with human *in vitro* data showing that gonadal differentiation is more efficient following lateral mesoderm specification (Knarston et al., 2020). The AGP becomes identifiable as a thickening of the CE on the ventral surface of the mesonephros at ~E9.5 in mouse and 4-5 post-conception weeks in humans (Carnegie stages 13-15) (Hatano et al., 1996; Ikeda et al., 1994; Sucheston and Cannon, 1968). It forms through CE proliferation and basement-membrane remodeling, allowing epithelial ingression and formation of a multilayered structure (Kusaka et al., 2010). CE progenitors of the AGP express key transcription factors, including WT1 (Sasaki et al., 2021; Wilhelm and Englert, 2002) and GATA4 (Hu et al., 2013), followed by LHX9 (Birk et al., 2000) and EMX2 (Miyamoto et al., 1997).

As the AGP gives rise to both adrenal and gonadal lineages, disruption of its formation or specification can impair development of both organs (Val and Swain, 2010). Targeted inactivation of critical regulators such as *Nr5a1, Wt1, Cbx2, Cited2, Lhx9 Pbx1, Odd1*, *Six1/4* or perturbation of IGF signaling, results in a range of adrenal and gonadal defects, including impaired genital ridge formation, gonadal dysgenesis or degeneration, and, in some cases, sex reversal(Birk et al., 2000; Fujimoto et al., 2013; Katoh-Fukui et al., 2005; Kreidberg et al., 1993; Luo et al., 1994; Pitetti et al., 2013; Schnabel et al., 2003; Val et al., 2007; Wang et al., 2005).

In both mouse and human, the adrenocortical lineages are specified in the anterior CE of the AGP, whereas the gonadal lineages emerge later in the posterior region (Cheng et al., 2022; Neirijnck et al., 2023). However, the spatial and temporal dynamics of primordium formation differ between species (**Figure 2**). In humans, the adrenal primordium is specified earlier in the anterior CE and is clearly separated from the gonadal primordium, which arises later in the caudal region (Cheng et al., 2022). In mouse, the AGP splits into adjacent cranial adrenal and caudal gonadal primordia that are specified at approximately the same time (Neirijnck et al., 2023). Anterior CE cells migrate and condense to form the adrenal primordium, while posterior CE cells, together with primordial germ cells, contribute to gonad formation. These structures become morphologically distinct by ~E11.0 in mouse (Hatano et al., 1996) and ~4-5 PCW in humans (Hanley et al., 1999; Hanley et al., 2001).

**Regionalization of CE progenitors along the anteroposterior axis controls adrenal and gonadal lineage specification**

Recent studies have clarified how the AGP segregates into two distinct primordia (Cheng et al., 2022; Neirijnck et al., 2023). In humans, anteroposterior patterning of the CE is associated with distinct HOX domains (**Figure 2**): the adrenocortical lineage preferentially expresses anterior *HOX* genes (*HOX1-7*), whereas gonadal lineages express more posterior *HOX* genes (*HOX8-13*)(Cheng et al., 2022). Similarly, early mouse adrenal progenitors show higher expression of anterior *Hox* paralogs, including *Hoxa2, Hoxb5,* and *Hoxc5-6* (Neirijnck et al., 2023). Thus, differential *Hox/HOX* expression may establish regional identity along the anteroposterior axis and thereby direct adrenal or gonadal fates. The distinct induction of these fates is also likely achieved through specific signaling pathways that modulate anterior-posterior regionalization. In mouse, canonical WNT signaling is repressed in uncommitted AGP cells and remains low in the adrenal lineage, whereas it is activated in gonadal progenitors. In parallel, cells of the AGP and the adrenal primordium activate Rho/planar cell polarity pathways, undergoing cytoskeletal remodeling, epithelial-to-mesenchymal transition, and interactions with extracellular matrix components.

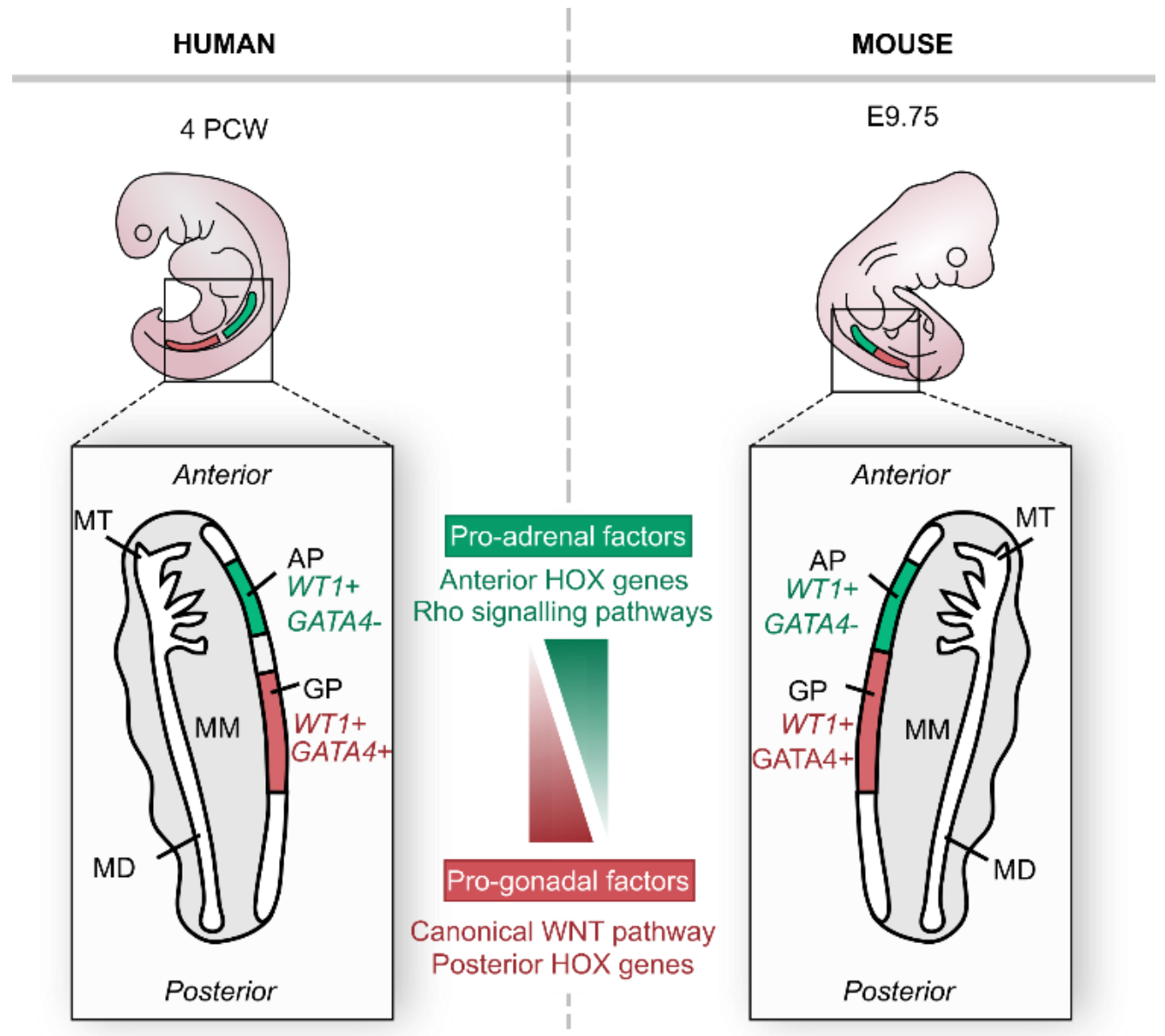

**Figure 2. Formation of adrenal and gonadal primordia along the anteroposterior axis.** Adrenocortical and gonadal lineages arise from coelomic epithelium progenitors in a spatially and temporally coordinated manner. In humans, the adrenal primordium (AP) forms earlier and is clearly separated from the gonadal primordium (GP), whereas in mice, both derive from a common adrenogonadal primordium that segregates into a cranial AP and a caudal GP. This regionalization is regulated by HOX genes and local signaling pathways, including Rho and canonical WNT signaling, which together direct adrenal versus gonadal fate. Abbreviations: AP, adrenal primordium; GP, gonadal primordium, MT, mesonephric tubules; MD, mesonephric duct. Adapted from (Wilhelm et al., 2025).

Overall, emerging evidence indicates that CE progenitors may already be biased toward adrenal or gonadal fate before ingression, suggesting that lineage specification begins within the epithelium itself (Neirijnck et al., 2023). This spatial pre-patterning, integrating regional transcriptional identity with local signaling, may also operate later within the gonadal primordium, contributing to distinct somatic lineages during ovarian and testicular development, as discussed below.

## Formation of the bipotential gonad from CE progenitors

The gonadal primordium, or genital ridge, emerges from the caudal part of the AGP at ~E11.0 in mouse and ~4-5 weeks of human embryonic development. Lineage-tracing studies have shown that, in both XY and XX gonads, cells delaminating from the CE are the major source of gonadal somatic cells (Karl and Capel, 1998; Schmahl et al., 2000). Consistently, transcriptomic trajectory analyses indicate that most supporting, supporting-like and steroidogenic progenitors derive from a common progenitor cell population (Garcia-Alonso et al., 2022; Lardenois et al., 2026; Mayere et al., 2022; Stevant et al., 2019; Stevant et al., 2018). These CE multipotent progenitors are characterized by the expression of key transcription factors, including GATA4, WT1 and NR5A1. Growth of the gonadal primordium relies on asymmetric divisions of CE progenitors, whereby some daughter cells remain in the epithelium, while others undergo epithelial-to-mesenchymal transition (EMT), and migrate into the developing gonad (Karl and Capel, 1998; Kusaka et al., 2010; Lin et al., 2017; Schmahl et al., 2000). By ~E10.5-E11.5, the underlying basement membrane is discontinuous, facilitating ingression through migration or oriented cell division. This process depends on proper epithelial polarity, regulated in part by the NOTCH antagonist NUMB, which is asymmetrically localized in CE cells and directs progenitor allocation toward gonadal somatic fates (Lin et al., 2017).

Reflecting their central role, CE-derived cells give rise to the three major gonadal somatic lineages: supporting, steroidogenic/stromal, and supporting-like cells (SLCs) (Ademi et al., 2022; Capel, 2000; DeFalco et al., 2011; Garcia-Alonso et al., 2022; Karl and Capel, 1998; Liu et al., 2016; Mayere et al., 2022; Mork et al., 2012; Schmahl and Capel, 2003; Stevant et al., 2019; Stevant and Nef, 2018). The supporting lineage gives rise to Sertoli cells in XY gonads and pre-granulosa cells in XX gonads, while the steroidogenic/stromal lineage forms Leydig and theca/stromal steroidogenic cells, respectively. SLCs are specified as early as E10.5 and contribute to the rete testis and rete ovarii (Kulibin and Malolina, 2020; Mayere et al., 2022; McKey et al., 2022).

In parallel, the developing gonad is colonized by migrating primordial germ cells (PGCs) and cells from the adjacent mesonephros (Buehr et al., 1993; DeFalco et al., 2011; McLaren, 2003; Molyneaux et al., 2001; Rotgers et al., 2018). As this Review focuses on somatic lineages, we refer readers to dedicated reviews of germ cell biology (Cooke and Moris, 2021; Hancock et al., 2021; Roelen and Chuva de Sousa Lopes, 2022; Saitou and Yamaji, 2012).

A central unresolved question is how multipotent CE progenitors become fate-restricted. Lineage bias could be established within the CE before ingression, after progenitors enter the gonad, or through a combination of both mechanisms. Defining when and where this restriction occurs, whether it differs between XX and XY gonads, and the molecular pathways controlling it is therefore critical. Recent studies are beginning to address these questions and reveal how early lineage specification shapes testicular and ovarian development.

## Origin of gonadal somatic lineages: early fate specification in CE Progenitors of human and mouse ovaries

A recent study reveals that CE progenitors overlying the developing ovary are transcriptionally heterogeneous and dynamically biased toward supporting or stromal/steroidogenic (SP) fates in both mouse and human (Djari et al., 2026) (**Figure 3**). All CE cells express the classical epithelial marker *Upk3b/UPK3B* but some subpopulation co-expressed either SP (*Nr2f2/NR2F2*, *Arx/ARX, Pdgfra/PDGFRA, Tcf21/TCF21*) or supporting (*Bmp2, Irx3/IRX3, Runx1, Wnt6/WNT6, LHX2*) markers and transcriptional signatures resembling their respective target lineages, consistent with fate priming. In both species, CE progenitors transition from an early SP-biased state to a transient phase in which supporting- and stromal/steroidogenic-biased populations coexist, to a predominantly supporting-biased state. CE fate bias is also spatially organized along the ovarian surface: SP-biased CE are enriched near the gonad-mesonephros interface whereas supporting-biased CE are predominantly located in the central region of the ovary. Importantly, intragonadal SP and pre-granulosa cells localize directly beneath their corresponding CE domains, supporting a model in which fate-primed CE cells delaminate locally to establish distinct somatic compartments.

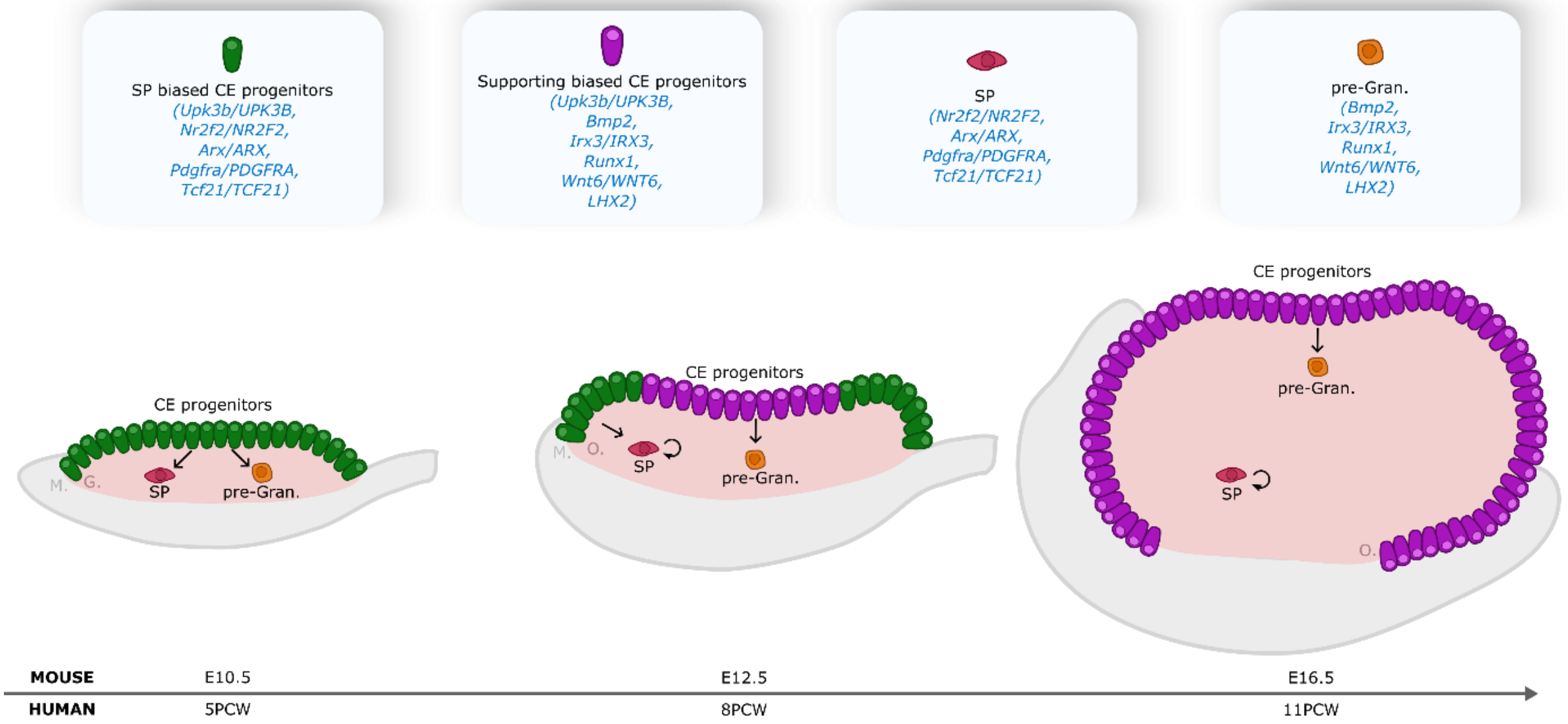

**Figure 3. Spatiotemporal priming of coelomic epithelium progenitors along the ovarian surface.** Coelomic epithelium (CE) progenitors of the developing ovary are transcriptionally heterogeneous and dynamically primed toward supporting or stromal/steroidogenic fates. In both human and mouse, CE priming progresses from an early stromal/steroidogenic-biased state, through a transient dual-primed phase, to a predominantly supporting-biased state. Spatially, CE fate-biased domains locally give rise to distinct ovarian lineages, establishing an early blueprint for somatic differentiation.

These findings are consistent with Bunce et al. (Bunce et al., 2023), who reported center-biased ingression of supporting progenitors in both XY and XX gonads, producing higher supporting cell density in the central domain.

Together, these data support a model in which spatial and temporal patterning of CE progenitors governs lineage allocation and the establishment of ovarian somatic architecture.

### Somatic cell differentiation during gonadal development in XX and XY gonads

The following sections address the process of gonadal sex determination and differentiation, during which the three principal somatic lineages adopt sex-specific fates (**Figure 4**). This includes the differentiation of supporting progenitors into Sertoli cells in the developing testis or pre-granulosa cells in the developing ovary; the steroidogenic/stromal lineage into fetal

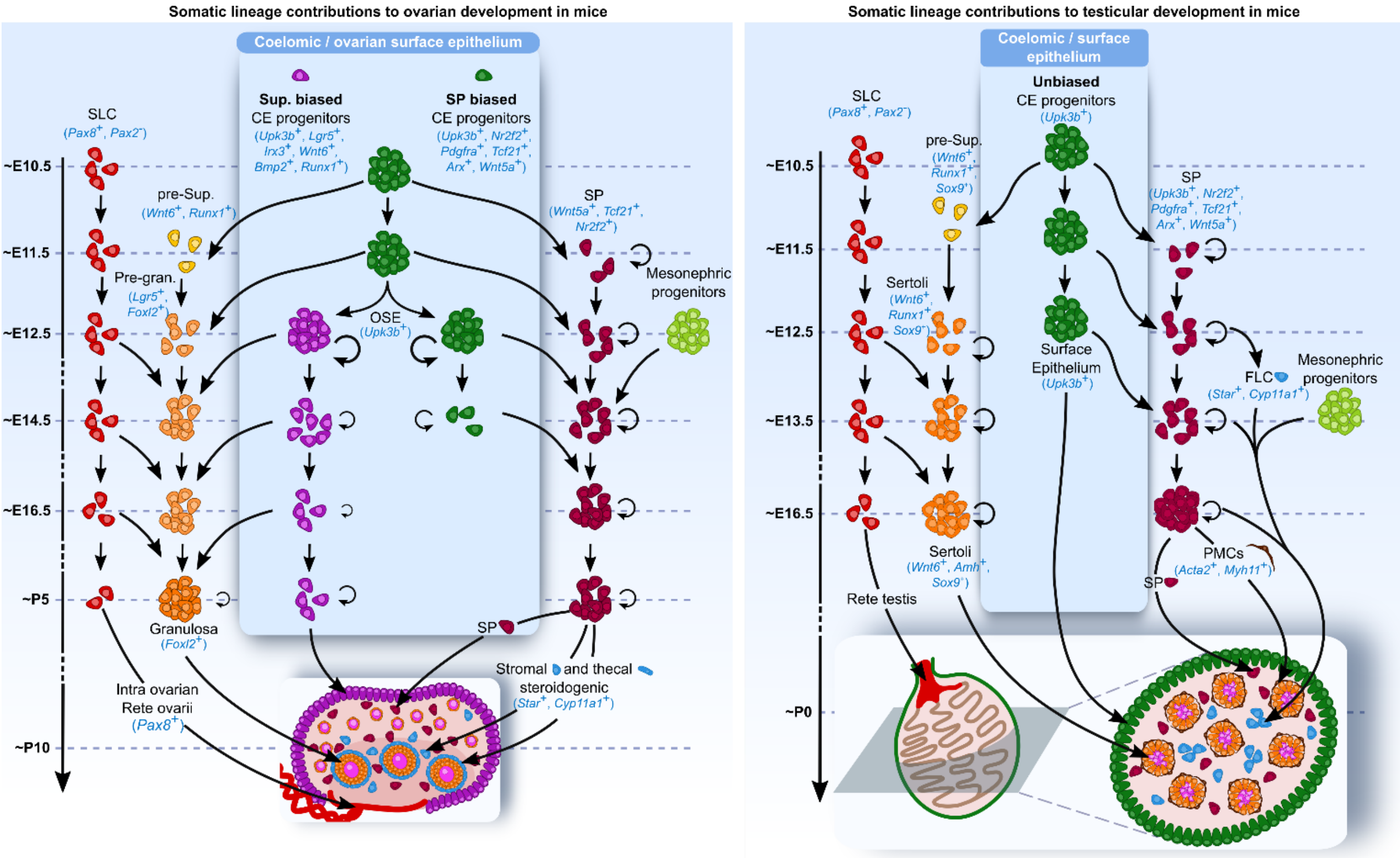

**Figure 4. Somatic cell differentiation during gonadal development in XX and XY gonads.** Contribution of the three principal somatic lineages to the formation of the testis (A) and ovary (B) during mouse development. Representative markers for each somatic cell population are indicated in blue. Abbreviations: CE, coelomic epithelium; SLCs, supporting-like cells; SPs, steroidogenic/stromal progenitors; IPs, interstitial steroidogenic progenitors; pre-gran., pre-granulosa cells; PMCs, peritubular myoid cells; FLCs, fetal Leydig cells.

Leydig cells or theca cells; and the supporting-like cell (SLC) lineage into the rete testis or rete ovarii.

**Emergence of the supporting cell lineage and differentiation into Sertoli or pre-granulosa cells**

*Supporting progenitors in the developing gonads*

Fate restriction of multipotent CE progenitors toward the supporting lineage precedes their sex-specific differentiation into Sertoli or pre-granulosa cells. In both XX and XY gonads, a subset of CE progenitors becomes restricted to the supporting lineage (Stevant et al., 2019; Stevant et al., 2018). Only subsequently do these supporting progenitors activate sex-specific genetic programs that drive their differentiation into Sertoli cells or pre-granulosa cells. Supporting-lineage specification occurs around E11.0-E12.0 and involves a similar transcriptional program in both sexes (Stevant et al., 2019; Stevant et al., 2018). This transient bipotential state includes expression of both pro-Sertoli genes, such as *Fgf9* and *Dmrt1*, and pre-granulosa-associated genes, including *Wnt4, Runx1,* and *Dax1* (Jameson et al., 2012b; Munger et al., 2013). Supporting progenitors therefore remain competent to adopt either fate. Consistently, ectopic *Sry* expression in XX supporting cells during the normal developmental window, is sufficient to induce Sertoli cell differentiation (Albrecht and Eicher, 2001; Harikae et al., 2013). By ~E11.5, supporting progenitors in XY gonads have already initiated differentiation into Sertoli cells (Hiramatsu et al., 2003), whereas in XX gonads this process occurs later, with supporting progenitors differentiating into pre-granulosa cells at subsequent stages (Stevant et al., 2019).

Despite these advances, the process of supporting-lineage specification remains poorly understood, particularly the relative contributions of intrinsic determinants and extrinsic signals from the gonadal microenvironment. Recent work suggests that supporting-lineage bias is already established within the CE of developing mouse and human ovaries (Djari et al., 2026). A subpopulation of supporting-primed CE cells has been identified, predominantly localized in the central ovarian surface epithelium. Whether similar priming occurs in XY gonads remains unclear. Because Sertoli differentiation begins around E11.5, approximately one day before pre-granulosa differentiation(Stevant et al., 2019), supporting-lineage restriction in XY gonads may involve a distinct, more rapid mechanism, potentially without prior CE priming..

*Supporting progenitors in the developing testis and Sertoli cell differentiation*

Pioneering experiments in the early 1990s using XY and XX mouse chimeras demonstrated that both XY and XX cells contribute to major testicular lineages (Palmer and Burgoyne, 1991). However, Sertoli cells were the only cell type to show a bias in favor of the XY component, indicating that their differentiation is driven by a cell-autonomous mechanism dependent on a testis-determining gene located on the Y chromosome within XY supporting progenitors. This gene was subsequently identified as *Sry* (sex-determining region on the chromosome Y) (Berta et al., 1990; Gubbay et al., 1990; Sinclair et al., 1990). *Sry* encodes a transcription factor belonging to the high mobility group (HMG)-box family, which binds to the minor groove of DNA and induces DNA bending, thereby regulating gene expression (Ferrari et al., 1992; Pontiggia et al., 1994; Pontiggia et al., 1995).

In mouse, *Sry* is expressed within a narrow window, from approximately E10.5 to E12.5, marking the onset of testis determination (Bullejos and Koopman, 2001; Hacker et al., 1995; Jeske et al., 1995; Sekido et al., 2004; Warr et al., 2012). Its expression was initially reported to spread from the gonadal center toward the poles before rapid downregulation (Bullejos and Koopman, 2001; Warr et al., 2012; Wilhelm et al., 2005), although this spatial pattern has recently been questioned (Bunce et al., 2023). This narrow temporal window is critical in mouse, as delays or reductions in *Sry* expression, for example in C57BL6/J mice carrying the $Y^{POS}$ chromosome, can impair testis determination (Bullejos and Koopman, 2005; Eicher et al., 1982; Hacker et al., 1995; Taketo et al., 1991). In humans, the *SRY* expression is restricted to supporting cell precursors but occurs over a broader, less sharply defined temporal window, beginning around 6-7 PCW (Garcia-Alonso et al., 2022; Hanley et al., 2000; O'Shaughnessy et al., 2007).

*Sry* is both necessary and sufficient for testis development (Gubbay et al., 1990; Kato et al., 2013; Koopman et al., 1991). Loss of *Sry* causes XY gonads to develop as ovaries (Lovell-Badge and Robertson, 1990). In humans, mutations in *SRY* account for approximately 10-20% of cases of 46,XY gonadal dysgenesis. These mutations frequently affect the HMG box domain, impairing DNA binding, nuclear localization, or transcriptional activity, and thereby preventing activation of *SOX9* and downstream testis-determining pathways (Cameron and Sinclair, 1997; Harley et al., 2003). Outside the HMG box, the SRY protein is highly divergent between species (Zhao and Koopman, 2012). Mouse SRY contains a glutamine-rich C-terminal domain absent from human SRY. Moreover, the mouse *Sry* gene comprises two exons, and only the two-exon isoform (SRY-T), rather than the canonical single-exon form (SRY-S), functions as the *bona fide* testis-determining factor (Miyawaki et al., 2020). Despite these sequence divergences, SRY function is conserved, as human *SRY* expressed under mouse regulatory control, as well as a goat *SRY* BAC transgene, can induce male development in XX mice (Lovell-Badge et al., 2002; Pannetier et al., 2006).

In XY embryos, *Sry* is expressed in supporting precursors, directly upregulating *Sox9* and thereby initiating their differentiation into pre-Sertoli cells (Koopman et al., 1991; Sekido et al., 2004; Sinclair et al., 1990). *SOX9*, an autosomal HMG-box transcription factor, subsequently acts as the master regulator of Sertoli cell differentiation, while the primary role of *Sry* is to activate *Sox9* within a narrow developmental window (Sekido et al., 2004; Sekido and Lovell-Badge, 2008; Wilhelm et al., 2005). Consistent with this, ectopic expression of *Sox9* induces Sertoli cell differentiation and testis development in XX embryos, whereas *Sox9* inactivation in XY gonads causes ovarian development (Barrionuevo et al., 2006; Chaboissier et al., 2004; Vidal et al., 2001).

In mice, *Sox9* is initially expressed at low levels in bipotential gonads but is rapidly upregulated in XY supporting cells following *Sry* activation while being actively suppressed in pre-granulosa cells of the ovary (Morais da Silva et al., 1996; Sekido et al., 2004; Stévant et al., 2025). Once SOX9 reaches a critical threshold, it downregulates *Sry* and sustains its own expression through positive feedback loops involving NR5A1, thereby stabilizing Sertoli identity (Gonen and Lovell-Badge, 2019; Sekido and Lovell-Badge, 2008). SOX9 also upregulates several downstream targets, including fibroblast growth factor 9 (*Fgf9*), anti-Müllerian hormone (*Amh*) and prostaglandin D2 synthase (*Ptgds*) (Arango et al., 2008; Kim et al., 2006; Moniot et al., 2009; Rahmoun et al., 2017; Wilhelm et al., 2007). In the absence of sufficient *SOX9* activation, supporting progenitors instead engage the ovarian pathway (Lavery et al., 2011).

Early models proposed that Sertoli differentiation initiates centrally, reflecting the center-to-pole expression of *Sry* and *Sox9*, and spreads toward the poles via paracrine signals such as FGF9 or prostaglandins (Bullejos and Koopman, 2001; Hiramatsu et al., 2010; Wilhelm et al., 2005). More recent data challenge this model, suggesting that *Sry* and *Sox9* are initially more broadly distributed across the gonad and that Sertoli cell fate does not propagate through a simple paracrine relay. Instead, a center-biased ingression of supporting precursors generates higher density centrally, potentially promoting and stabilizing Sertoli differentiation through local density-dependent mechanisms (Bunce et al., 2023).

Testis development requires a critical number of *Sox9*-positive supporting cells. In XY-XX chimeras, an XY contribution below ~20%, leads to ovotestes or ovarian development (Palmer and Burgoyne, 1991). Several reinforcing mechanisms act to expand and maintain the number of *Sox9*-positive pre-Sertoli cells in the developing testis. XY cells can recruit neighboring XX cells to a Sertoli fate through prostaglandin D2 (PGD2) signaling, which induces *Sox9* expression in adjacent cells and establishes a positive feedback loop via *Ptgds* (Moniot et al., 2009; Wilhelm et al., 2005). In parallel, FGF9 signaling contributes to sustaining *Sox9* expression and promoting testis development, although its primary role may be repression of pro-ovarian pathways such as *Wnt4* (Jameson et al., 2012a; Kim et al., 2006). Together, these mechanisms ensure robust amplification and maintenance of *Sox9* expression, allowing testis differentiation to proceed despite the transient expression of *Sry*.

*Aggregation of pre-Sertoli cells into primitive testis cords and testis cord partitioning*

After *Sox9* activation, pre-Sertoli cells aggregate around germ cells to form testis cords, at which point they are referred to as Sertoli cells. Testis cord formation establishes two functionally distinct compartments: the seminiferous cords/tubules, which support spermatogenesis, and the interstitial compartment, responsible for androgen and INSL3 production. In addition, cord formation plays a critical role in germ cell regulation by physically isolating gonocytes and preventing premature entry into meiosis. Although Sertoli cells express CYP26B1, which was initially proposed to block meiosis by degrading retinoic acid and suppressing STRA8 expression (Bowles et al., 2006; MacLean et al., 2007), recent evidence suggests that CYP26B1 may instead metabolize an unidentified meiosis-inducing factor or by contributing to the production of an unknown meiosis-preventing signal (Bellutti et al., 2019; Kumar et al., 2011; Perrotta et al., 2025).

Despite its importance, testis cord assembly remains poorly understood. Cord formation occurs normally in the absence of germ cells, demonstrating that germ cells are dispensable for cord initiation (Escalante-Alcalde and Merchant-Larios, 1992; McCoshen, 1982, 1983; Merchant, 1975). Testis cord assembly is therefore primarily driven by Sertoli-Sertoli interactions rather than aggregation around germ cells.

In contrast to germ cells, vascular endothelial cells have been shown to play an important role in testis cord formation (Combes et al., 2009b; Cool et al., 2011; Coveney et al., 2008). At E11.5, the mesonephros adjacent to the gonad contains a dense microvascular plexus composed of an extensive network of microvessels. Following activation of the testis pathway by *Sry*, this mesonephric vascular plexus undergoes rapid remodeling and breakdown, releasing individual endothelial cells that delaminate and migrate into the gonad, where they contribute to the developing vasculature. These cells reorganize to form, the primitive coelomic vessel by E12.5. Importantly, endothelial cell migration partitions the gonad into approximately ten avascular domains, where testis cord form over the subsequent ~12 hours (Coveney et al., 2008). Disruption of endothelial cell migration impairs the development of testis cord formation, highlighting their essential role in early gonadal patterning (Combes et al., 2009b; Coveney et al., 2008). As development proceeds, Sertoli cells proliferate, reinforce cell-cell contacts, and drive cord elongation and remodeling into structures of more uniform diameter (Nel-Themaat et al., 2011). By ~E13.5, cords acquire an epithelial organization and become surrounded by peritubular myoid cells and extracellular matrix, before elongating, bending, and coiling to generate the complex architecture of the adult testis (Combes et al., 2009a; Nel-Themaat et al., 2009).

Disruption of Hedgehog signaling further highlights the importance of interactions between Sertoli cells and interstitial lineages in testis cord morphogenesis. In *Dhh* or *Hhat* mutant mice, defective fetal Leydig and peritubular myoid cell specification is associated with fewer and morphologically abnormal cords(Callier et al., 2014; Clark et al., 2000; Pierucci-Alves et al., 2001). Nevertheless, major mechanistic questions remain: how pre-Sertoli cells self-organize into elongated cords rather than simple aggregates, how they progressively envelop germ cells during cord formation. In addition, the mechanisms by which nascent cords connect to the rete testis remain unknown, including how rete-testis cells coordinate with Sertoli cells to anchor and integrate each cord into a continuous tubular network.

*Supporting progenitors in the developing ovary and pre-granulosa cell differentiation*

The follicle is the functional unit of the ovary, supporting oocyte development and culminating in the release of a mature oocyte. Primordial follicle formation begins around PCW13 in humans, whereas in mice it occurs postnatally (Maheshwari and Fowler, 2008; Pelosi et al., 2015; Pepling, 2012). During folliculogenesis, pre-granulosa cells mature into granulosa cells as the primordial follicles transition into primary follicles (Hirshfield, 1991).

Pre-granulosa cells are generated in two asynchronous waves (Mork et al., 2012; Niu and Spradling, 2020; Zheng et al., 2014) (**Figure 5A**). The first wave arises around the time of gonadal sex determination and consists of early FOXL2-expressing cells that associate with germ cell cysts by E12.5 (Pisarska et al., 2011). These cells give rise to granulosa cells of medullary follicles (Mork et al., 2012; Zheng et al., 2014). The second wave originates from *Lgr5*$^{+}$ progenitors in the ovarian surface epithelium from approximately E14.5 until one week after birth (Mork et al., 2012; Ng et al., 2014; Rastetter et al., 2014). These cells downregulate *Lgr5,* and upregulate *Foxl2* postnatally, giving rise to cortical follicles that constitute the long-term follicular reserve (Cai et al., 2020; Feng et al., 2016; Ng et al., 2014; Niu and Spradling, 2020; Rastetter et al., 2014; Stevant et al., 2019).

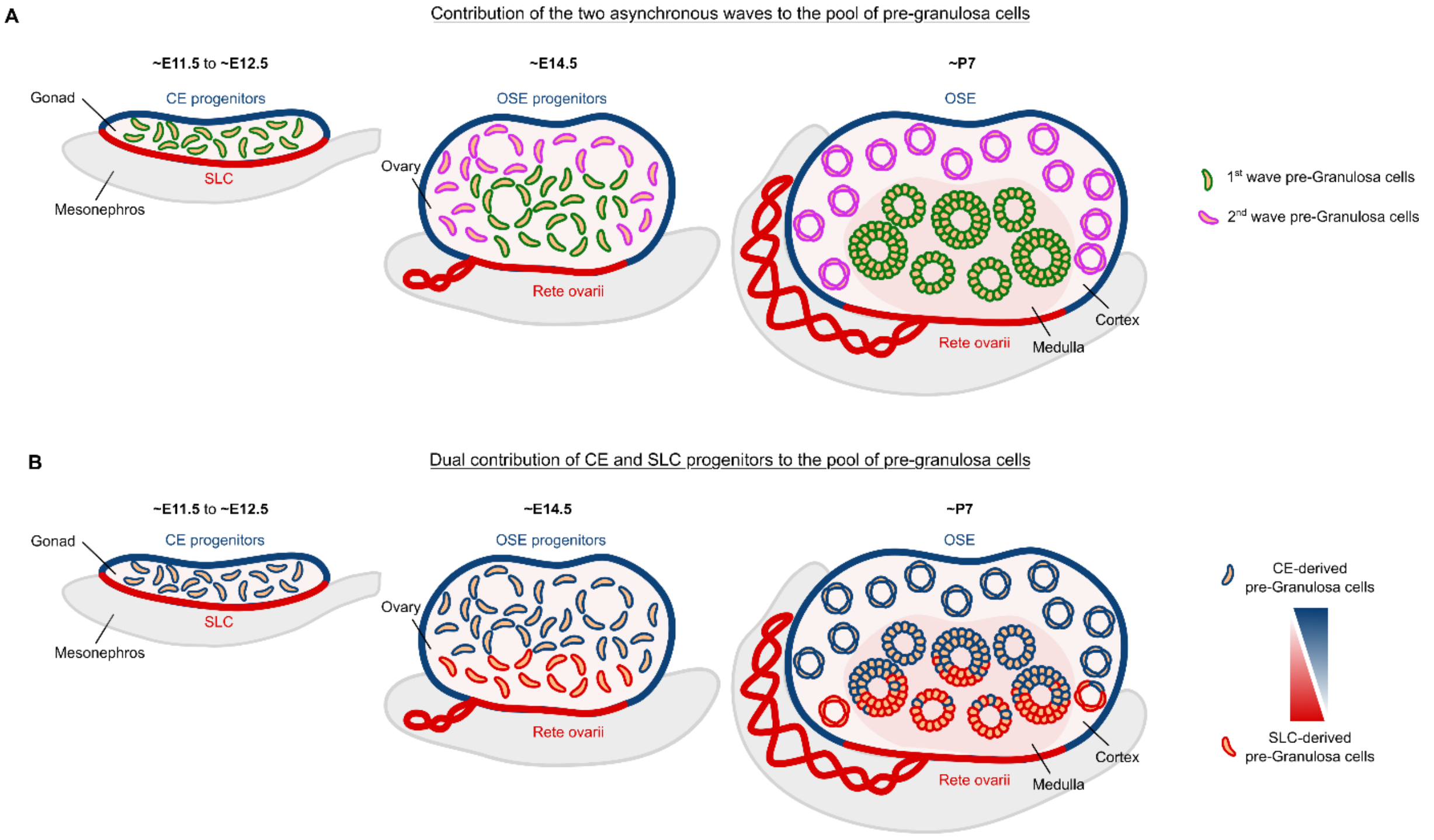

**Figure 5. Emergence of granulosa cells during mouse ovarian development.** (A) Two asynchronous waves of pre-granulosa specification contribute to the medullary (1st wave, ~E11.5 to ~E12.5) and cortical (2nd wave, ~E14.5 to ~P7) follicular compartments. (B) CE and SLC progenitors converge toward a common pre-granulosa fate, each contributing to both the medullary and cortical compartments. SLC-derived pre-granulosa cells are enriched near the rete ovarii, while CE-derived pre-granulosa cells predominate near the ovarian surface.
Abbreviations: CE, coelomic epithelium; OSE, ovarian surface epithelium; SLC, supporting-like cells.

Recent work has refined the functional significance of these two waves in mice (Yin and Spradling, 2025). Most first-wave follicles undergo atresia before puberty and contribute little, if at all, to fertility. After reaching preantral stages and forming theca layers, most lose their oocytes and granulosa cells, while retaining theca cells that contribute to the medullary steroidogenic compartment. Early fertility is largely supported by "boundary" or "wave 1.5" follicles located at the medullary-cortical interface. These follicles are recruited from around two weeks of age and generate most of the earliest mature oocytes, whereas cortical second-wave follicles are activated later, around puberty. A similar process may occur in humans, where follicles develop long before they could contribute to fertility. Folliculogenesis begins during fetal life, with 4-6% of follicles reaching the primary stage before birth. However, the developmental significance of these fetal and juvenile human follicles remains poorly understood.

In XX gonads, lineage-tracing and transcriptomic studies identify the CE as the primary source of pre-granulosa cells (Albrecht and Eicher, 2001; Karl and Capel, 1998; Mayere et al., 2022; Stevant et al., 2019). Recent work indicates that this process may involve fate priming within spatially organized CE domains in both mouse and human ovaries (Djari et al., 2026). Supporting-primed CE progenitors are enriched centrally and locally give rise to underlying pre-granulosa cells, as discussed above. Interestingly, the timing of these two waves of pre-granulosa emergence coincides with a temporal shift in CE identity from an SP-biased state to a supporting-biased state. This raises the possibility that first-wave pre-granulosa cells originate from SP-biased CE progenitors that retain the ability to delaminate and commit to the supporting lineage at early developmental stages, while second-wave pre-granulosa cells arise from the later supporting-biased CE population.

Finally, the CE is not the sole source of pre-granulosa cells. Supporting-like cells (SLCs) also contribute to the pre-granulosa lineage (Djari et al., 2026; Mayere et al., 2022). Lineage-tracing studies indicate that SLC-derived $FOXL2^+$ cells account for up to ~22% of pre-granulosa cells at E16.5, decreasing to ~10% in adulthood. By postnatal day 5 (P5), these cells are distributed in both medullary and cortical regions and display a gradient from the hilum toward the ovarian center, indicating a spatially organized contribution.

*Wt1-KTS isoform as the ovarian determining gene*

Recently, the alternatively spliced *WT1* isoform containing the lysine-threonine-serine (KTS) insertion, WT1-KTS, has emerged as a key determinant of ovarian development. In fact, the WT1-KTS isoform has all the characteristics of the long sought ovarian-determining factor: it is required for ovarian differentiation while its ectopic expression prevents testis development. In mice, *Wt1-KTS* drives supporting progenitors toward a pre-granulosa cell fate, acting upstream of pro-ovarian pathways such as RSPO1-WNT/β-catenin signaling (Gregoire et al., 2023; Motamedi et al., 2014). Loss of *Wt1-KTS* disrupts pre-granulosa cell differentiation and impairs ovarian development, whereas ectopic expression in XY gonads induces male-to-female sex reversal (Gregoire et al., 2023; Hammes et al., 2001). In humans, mutations affecting *WT1* splicing underlie Frasier syndrome, a 46,XY DSD characterized by gonadal dysgenesis (Barbaux et al., 1997; Klamt et al., 1998). Together, these findings identify WT1-KTS as a critical regulator of ovarian fate, functionally analogous to SRY in testis determination. However, important questions remain regarding the regulation and function of *WT1/Wt1* alternative splicing during early gonadal development. In particular, it remains unclear whether the relative abundance of *WT1/Wt1* isoforms varies with the priming state of CE progenitors and whether alternative splicing contributes to their commitment toward the supporting lineage.

WT1-KTS promotes RSPO1/WNT/β-catenin signaling and FOXL2 expression, two central regulators of ovarian development that control pre-granulosa cell differentiation and the maintenance of ovarian identity (Chassot et al., 2008; Liu et al., 2009; Migale et al., 2024; Uhlenhaut et al., 2009; Vainio et al., 1999). Canonical WNT signaling is tightly controlled at the receptor level by the opposing activities of ZNRF3/RNF43 and R-spondins. The E3 ubiquitin ligases ZNRF3 and RNF43 negatively regulate WNT signaling by promoting ubiquitination and degradation of the WNT receptor complex (de Lau et al., 2014; Hao et al., 2012; Koo et al., 2012). Conversely, R-spondins bind LGR4–6 receptors and promote ZNRF3/RNF43 internalization, thereby stabilizing WNT receptors and sustaining β-catenin signaling (de Lau et al., 2011). *Rspo1* and *Wnt4* are direct WT1 targets in the developing kidney, suggesting that WT1 may similarly promote their expression in XX gonads (Essafi et al., 2011; Motamedi et al., 2014). In XY gonads, ectopic activation of WNT/β-catenin signaling disrupts normal testis development (Maatouk et al., 2008). Conversely, ZNRF3-mediated repression of WNT signaling limits pathway activity and contribute to the appropriate balance of progenitor cell fate decisions. Accordingly, perturbation of ZNRF3 or inappropriate activation of WNT signaling alters supporting-cell specification and testis development (Gregoire et al., 2018; Harris et al., 2018; Jeays-Ward et al., 2004; Kay et al., 2024).

Canonical WNT signaling is also essential for establishing and maintaining ovarian identity, preventing premature differentiation of pre-granulosa cells and their subsequent transdifferentiation toward a Sertoli-like fate (Chassot et al., 2008; Koizumi et al., 2015; Lavery et al., 2012; Maatouk et al., 2013; Nicol and Yao, 2015; Richardson et al., 2020; Tomizuka et al., 2008; Vainio et al., 1999). Disruption of this pathway can result in ovotestis formation, consistent with 46,XX DSD associated with mutations in *WNT4* or *RSPO1* (Mandel et al., 2008; Parma et al., 2006; Tomaselli et al., 2008).

In mice, *Rspo1* expression declines postnatally, as FOXL2 becomes increasingly critical for maintaining ovarian identity and supporting folliculogenesis (De Cian et al., 2017; Ottolenghi et al., 2005; Pailhoux et al., 2002; Schmidt et al., 2004). FOXL2 is required for the squamous-to-cuboidal transition of granulosa cells during primary follicle formation and, together with TRIM28, maintains granulosa cell identity (Ottolenghi et al., 2005; Rossitto et al., 2022; Schmidt et al., 2004; Uhlenhaut et al., 2009). FOXL2 also regulates broad transcriptional programs involved in follicle formation and steroidogenesis (Migale et al., 2024). In goats, loss of FOXL2 leads to female-to-male sex reversal, highlighting its essential role in establishing and maintaining ovarian identity (Boulanger et al., 2014).

Overall, WT1-KTS, RSPO1/WNT/β-catenin signaling, and FOXL2 form a central regulatory axis that establishes and maintains ovarian identity (Imaimatsu et al., 2022; Suzuki et al., 2015). However, the mechanisms by which these pathways establish a stable ovarian developmental program remain incompletely understood.

**Emergence of the steroidogenic lineage and differentiation of fetal Leydig cells and theca cells**

Androgen synthesis is essential for sexual development and fertility in both sexes. In males, androgens drive sexual development and support spermatogenesis, whereas in females they contribute to follicular development and serve as precursors for estradiol synthesis (Franks and Hardy, 2018; Li et al., 2024; Walters and Handelsman, 2018). Androgens are produced primarily by interstitial Leydig cells in the testis and theca cells in the ovary. Despite their central role, the origin and mechanisms governing steroidogenic progenitor specification and differentiation remain incompletely understood

An additional key aspect to consider when comparing steroidogenesis in the developing testis and ovary is their distinct temporal dynamics, with testis differentiating earlier and more rapidly (Nicol et al., 2022; Stevant and Nef, 2019; Wilhelm et al., 2013). Fetal Leydig cells initiate androgens secretion shortly after testis differentiation, around E12.5 in mouse and 6-7 PCW in human. In contrast, ovarian theca cells arise much later and acquire full steroidogenic activity only after birth, most prominently at puberty, when they become LH-responsive during folliculogenesis (Barsoum and Yao, 2010; Fortune and Armstrong, 1977; Magoffin, 2005; Mannan and O'Shaughnessy, 1991; Richards et al., 2018). The theca layer comprises androgen-producing theca interna cells and structurally supportive theca externa cells (Magoffin, 2005). In humans, the balance between structural and androgenic cell fates is modulated by the reciprocal activity of TCF21 and NR5A1 (Guahmich et al., 2023).

*Steroidogenic progenitors in the developing testis and Leydig cells differentiation*

In the mouse testis, fetal Leydig cells (FLCs) arise around E12.5, shortly after Sertoli specification (Barsoum and Yao, 2010). Their numbers increase substantially between E12.5 and E16.5, then decline during the early postnatal period as they are replaced by adult Leydig cells (ALCs) (Chen et al., 2017; Teerds and Huhtaniemi, 2015). As FLCs and ALCs are largely non-proliferative (Byskov, 1986; Migrenne et al., 2001; Orth, 1982), their expansion relies on the recruitment and differentiation of steroidogenic progenitors (SPs) located in the interstitial compartment (Estermann et al., 2025). These progenitors constitute a major cell population in the fetal testis, exceeding 50% of total cells by E16.5 (Ademi et al., 2022) and express canonical markers, including *Wnt5a, Arx, Nr2f2, Tcf21* and *Pdgfra* (Ademi et al., 2022; Brennan et al., 2003; Cui et al., 2004; Eliveld et al., 2019; Estermann et al., 2025; Ge et al., 2006; Gui and Yuan, 2021; Kilcoyne et al., 2014; Miyabayashi et al., 2013; Perea-Gomez et al., 2025; Qin et al., 2008).

Androgen-producing cells in testis and ovary arise from multiple sources (DeFalco et al., 2011; Karl and Capel, 1998; Kumar and DeFalco, 2018; Liu et al., 2015; Liu et al., 2016; Rotgers et al., 2018). In the mouse testis, *in vivo* lineage tracing and single-cell transcriptomic analyses indicate that SPs originate from both CE-derived and mesonephros-derived populations (Ademi et al., 2022; Kumar and DeFalco, 2018). Early FLCs are predominantly CE-derived, mesonephric progenitors contribute increasingly only later in development (Kumar and DeFalco, 2018). These progenitors are multipotent, and their fate is temporally regulated: early progenitors give rise to FLCs, whereas later progenitors contribute to peritubular myoid cells (PMCs) and, postnatally, to ALCs (Ademi et al., 2022).

Differentiation of SPs into FLCs is promoted by Sertoli cell-derived signals, including DHH and PDGFRA pathways (Barsoum and Yao, 2011; Brennan et al., 2003; Schmahl et al., 2008; Yao et al., 2002), and is negatively regulated by NOTCH ligands from vascular and perivascular cells (Defalco et al., 2013; Kumar and DeFalco, 2018; Liu et al., 2016; Tang et al., 2008). Interestingly, Sertoli-derived signals become dispensable after E14.5 for FLC differentiation and maintenance (Souali-Crespo et al., 2023). In addition to paracrine inputs from neighbouring cells, the differentiation of FLCs is regulated by the cell-autonomous activity of transcription factors such as ARX, TCF21, PBX1, NR2F2, MAF and MAFB in SP, although their precise roles remain unclear (Ademi et al., 2022; Cui et al., 2004; Estermann et al., 2025; Li et al., 2021; Miyabayashi et al., 2013; Perea-Gomez et al., 2025; Schnabel et al., 2003). Despite their abundance, SPs remain poorly characterized, particularly regarding why some progenitors differentiate early while others remain quiescent and only later give rise to ALCs.

A recent mouse model enabling specific ablation of ALCs revealed the presence of persistent SPs capable of regenerating a fully functional ALC population (Boudreau et al., 2026), consistent with earlier findings in ethane dimethanesulfonate (EDS)-treated rats (Molenaar et al., 1985; Teerds et al., 1990). Following ALC ablation, NR2F2-positive cells, likely representing persistent SPs, remained adjacent to the peritubular myoid cell (PMC) layer. Some subsequently acquired TdTomato expression, suggesting their differentiation into Leydig cells. These findings implicate the peritubular compartment as a potential regenerative niche containing persistent SPs, although lineage tracing is required to confirm the contribution of SPs. The local signals driving ALC regeneration remain unclear, but PDGF and Hedgehog signaling are promising candidates, as in FLCs development (Chen et al., 2020; Wang et al., 2026).

*Steroidogenic progenitors in the developing testis and peritubular myoid cells (PMCs)*

Peritubular myoid cells (PMCs) are specialized smooth muscle-like cells that surround the testis cords and seminiferous tubules (Maekawa et al., 1996; Mayerhofer, 2013). In rodents, PMCs form a single layer, whereas in humans they are organized into multiple layers within the lamina propria (Maekawa et al., 1996). PMCs exhibit contractile properties and contribute to the transport of spermatozoa and tubular fluid.

PMCs also contribute to testis cord morphogenesis by forming a contractile sheath around the seminiferous epithelium and by depositing extracellular matrix (ECM) in cooperation with Sertoli cells (Skinner et al., 1985; Tung et al., 1984). They are critical for basement membrane formation and for maintaining cord architecture. During fetal development, PMCs emerge shortly after testis cord formation and are morphologically identifiable around E13.5 in mice and at approximately 12 weeks of gestation in humans (Jeanes et al., 2005; Ostrer et al., 2008). Notably, PMCs are specific to the testis, with no direct counterpart in the ovary.

Recent studies indicate that PMCs arise from SPs, which also give rise to fetal and adult Leydig cells (Ademi et al., 2022). Early SPs (E12.5-E13.5) preferentially differentiate into fetal Leydig cells, whereas later SPs progressively lose this capacity and instead acquire the potential to differentiate into PMCs by around E16.5. This plasticity is lost after birth. In addition, transcriptional analyses suggest the existence of at least two distinct PMC subpopulations in the mouse testis (Ademi et al., 2022). Lineage-tracing studies using SP markers such as *Nestin* and *Tcf21* further support a common origin of Leydig cells and PMCs from interstitial progenitors (Kumar and DeFalco, 2018; Shen et al., 2021).

Few animal models exhibit early defects in PMC development, but evidence highlights a key role for Sertoli cells in their specification. In particular, Sertoli cell-derived DHH signaling has emerged as a central regulator of both PMC and Leydig cell differentiation (Pierucci-Alves et al., 2001; Yao et al., 2002). DHH acts through the PTCH1 receptor, which is expressed in SPs at the time of their differentiation (Yao et al., 2002). Disruption of Hedgehog signaling, as observed in *Dhh* or *Hhat* mutant mice, results in defects in both PMC and Leydig cell development, highlighting Sertoli-interstitial crosstalk (Callier et al., 2014; Clark et al., 2000; Pierucci-Alves et al., 2001; Yao and Capel, 2002). However, the mechanisms governing SP allocation to Leydig versus PMC lineages remain poorly understood.

*Steroidogenic progenitors in the developing ovary and theca cells differentiation*

In the mouse ovary, theca cells first appear around one week after birth, in secondary follicles. This timing coincides with the acquisition of luteinizing hormone (LH) responsiveness and the activation of steroidogenic enzymes (Magoffin and Weitsman, 1994). Theca cells produce androgens in response to LH stimulation, which are subsequently converted into estrogens by follicle-stimulating hormone (FSH)-induced aromatase in the adjacent granulosa cells (Edson et al., 2009).

As in the testis, theca cells arise from multiple sources, including a major source from intrinsic ovarian progenitors (*Wt1*$^{+}$ cells) including *Nr2f2*$^{+}$ stromal cells and *Gli1*$^{+}$ mesenchymal cells migrating from the mesonephros (Brown et al., 2025; Liu et al., 2015). Recent work further indicates that the stromal/steroidogenic lineage is already primed within the CE of developing mouse and human ovaries (Djari et al., 2026). As described above, SP-biased CE progenitors are initially widespread but become progressively restricted toward the gonad-mesonephros interface, where they locally delaminate and contributes directly to intraovarian steroidogenic progenitors (SPs) (Djari et al., 2026).

SPs proliferate extensively and account for approximately half of all fetal ovarian cells in both human and mouse, mirroring the expansion of SPs observed in the developing testis (Ademi et al., 2022; Djari et al., 2026). In both humans and mice, ovarian SPs express canonical markers such as *Arx/ARX, Pdgfra/PDGFRA, Wnt5a, Nr2f2* and *Tcf21/TCF21*, which are likewise characteristic of testicular SPs (Brennan et al., 2003; Cui et al., 2004; Eliveld et al., 2019; Estermann et al., 2025; Ge et al., 2006; Kilcoyne et al., 2014; Miyabayashi et al., 2013; Perea-Gomez et al., 2025; Qin et al., 2008; Shen et al., 2021). Lineage tracing studies further demonstrate that fetal *Wnt5a*$^{+}$ SPs persist and give rise to steroidogenic theca and stromal cells in the ovary, as well as fetal Leydig cells in the testis (Ademi et al., 2022; Brown et al., 2025; Estermann et al., 2025).
Differentiation of SPs into theca cells is initiated by the expression of the steroidogenic lineage marker *Gli1*, induced by paracrine signaling from granulosa cell-derived factors Desert hedgehog (DHH) and Indian hedgehog (IHH)(Cowan and Quirk, 2021; Richards et al., 2017). Ovaries deficient in *Dhh/Ihh* exhibit loss of theca layers, reduced steroid production, arrested folliculogenesis, and failure to form corpora lutea (Liu et al., 2015). Granulosa-cell DHH/IHH production is itself regulated by oocyte-derived Growth differentiation factor 9 (GDF9) (Liu et al., 2015). This highlights the critical interplay among oocytes, granulosa cells, and theca cells in the formation of functional theca layers, which are critical for proper follicular development and ovarian function (Liu et al., 2015).

**Emergence of the Supporting-like cell (SLC) lineage and its contribution to rete structures, Sertoli, and granulosa cells**
Beyond its established role in generating supporting and steroidogenic lineages, the coelomic epithelium (CE) also gives rise to a population of supporting-like cells (SLCs) in both mouse and human (Garcia-Alonso et al., 2022; Kulibin and Malolina, 2020; Lardenois et al., 2026; Mayere et al., 2022; Taelman et al., 2024). These cells are referred to as SLCs due to their transcriptional similarity to the supporting lineage and the expression of the transcription factor *Pax8*. SLCs are specified as early as E10.5, prior to sex determination, making them among the earliest somatic lineages specified in the bipotential gonad. At this stage, a subset of CE cells already express *Pax8*, indicating early commitment toward the SLC fate (Mayere et al., 2022). SLCs localize along the gonad-mesonephros interface in both XY and XX gonads and express canonical gonadal markers (*Nr5a1, Gata4, Wt1*), while also co-expressing pro-testicular (*Sox9*) and pro-ovarian (*Wnt4*) genes, consistent with an initially sexually undifferentiated state. From E12.5 onward, SLCs progressively acquire Sertoli-like or pre-granulosa-like identities in developing testis and ovary, respectively. Lineage-tracing experiments further demonstrate that *Pax8*$^{+}$ progenitors contribute predominantly to the rete testis and rete ovarii, while also giving rise to a subset of Sertoli and pre-granulosa cells (Djari et al., 2026; Mayere et al., 2022).
In the adult testis, the rete testis forms an interconnected network of epithelial tubules located in the mediastinum, collecting luminal fluid from the seminiferous tubules and channeling it toward the efferent ducts, epididymis and vas deferens (Major et al., 2021). Defects in rete testis formation can impair sperm transport and lead to obstructive azoospermia and infertility. Although its fetal development remains incompletely understood, in the mouse, connections between testis cords, the rete testis, and mesonephric tubules are established by approximately E15.5 (Nakata and Iseki, 2019). In both mouse and human, rete testis cells display a dual molecular identity, expressing both supporting cell markers (*Sox9, Gata4, Wt1, Nr5a1*) and rete-associated markers (*Pax8, Krt8, Cdh1*) (Magers et al., 2016; Malolina and Kulibin, 2019; Omotehara et al., 2020).
The rete ovarii, the female counterpart of the rete testis, consists of a network of tubules located in the ovarian hilus and extending into the medulla (McKey et al., 2022; Wenzel and Odend'hal, 1985). It comprises three distinct regions that persist into adulthood: the intraovarian rete (IOR), extraovarian rete (EOR), and connecting rete (CR). All three regions express PAX8, but only the IOR appears to derive from the SLC lineage, whereas the EOR and CR originate from mesonephric tubules. Although its physiological role remains unclear, the rete ovarii has been proposed to contribute to initiation of meiosis and to serve as a source of pre-granulosa cells during embryonic development (Byskov, 1974; Smith et al., 2014). More recent work suggests that it may also function as a secretory structure, with luminal flow toward the ovary, potentially contributing to ovarian development and homeostasis (Anbarci et al., 2025).

A particularly notable feature of the SLC lineage is its ability to contribute not only to rete structures, but also to a subpopulation of supporting cells, including Sertoli cells and granulosa cells.. In the mouse testis, lineage-tracing experiments show that SLCs contribute to approximately 10% of the total Sertoli cell population at birth, with a higher density near the rete testis (Mayere et al., 2022). Similarly, in the developing ovary, SLCs also contribute to pre-granulosa cell population (Mayere et al., 2022) (**Figure 5B**). Integrated single-cell transcriptomics, spatial mapping, and lineage tracing analyses show that by E12.5, ~5% of pre-granulosa cells are SLC-derived. This proportion increases to ~22% by E16.5, with SLC-derived cells enriched near the rete ovarii and hilum, and declines to ~10% in adulthood (Djari et al., 2026). These findings demonstrate that granulosa cells arise from two distinct progenitor pools: CE-derived progenitors, which generate the majority of cortical granulosa cells forming the long-lived follicle reserve, and SLCs, which contribute a smaller, spatially restricted subpopulation. Whether these distinct developmental origins confer functional differences on granulosa cells remains unknown..

**Conclusion, remaining gaps and outline future directions**

Our understanding of mammalian gonadal development is shifting from a model centered primarily on sex-specific differentiation toward one in which progenitor competence, lineage priming, spatial organization, and fate restriction precede and shape the response to sex-determining signals. Recent studies indicate that the ovarian CE is not a homogeneous progenitor compartment but contains transcriptionally distinct populations biased toward supporting, steroidogenic/stromal, and SLC fates. These findings raise the broader question of how initially multipotent progenitors acquire, maintain, and ultimately resolve lineage biases during organogenesis.

Addressing these questions will require integration of spatial transcriptomics, lineage tracing, functional perturbation, epigenomic profiling, *in vitro* models, and computational modelling. Applying these approaches across developmental stages and species, particularly mouse and human, should reveal which principles of progenitor specification are conserved and which are species-specific. Beyond refining models of gonadal development, this framework should help explain how disruption of progenitor specification, lineage allocation, or differentiation contributes to DSD and infertility, while guiding the development of more faithful *in vitro* models of human gonad development.

**Acknowledgments**

We thank Aitana Perea-Gomez, Chloé Mayère, and Cyril Djari for their critical reading of the manuscript and valuable comments and suggestions. We also acknowledge the many researchers whose contributions have advanced our understanding of gonadal development. Given the breadth of the field and space limitations, we were unable to cite all relevant studies and apologize to colleagues whose important work could not be included.

This work was supported by the Swiss National Science Foundation (SNSF; grants 31003A_173070, 310030_200316, and 10.001.251 to S.N.); the Département de l'Instruction Publique of the State of Geneva (S.N.); the Agence Nationale de la Recherche (ANR; grants ANR-23-CE14-0012 HeteroSex and ANR-24-CE14-2476 Anaform to M.-C.C.); and the European Research Council (ERC; Advanced Grant ERC-2024-ADG-101199244, TESorOVA, to M.-C.C.).

.

**References**

Ademi, H., Djari, C., Mayere, C., Neirijnck, Y., Sararols, P., Rands, C.M., Stevant, I., Conne, B., and Nef, S. (2022). Deciphering the origins and fates of steroidogenic lineages in the mouse testis. Cell Rep *39*, 110935.
Albrecht, K.H., and Eicher, E.M. (2001). Evidence that Sry is expressed in pre-Sertoli cells and Sertoli and granulosa cells have a common precursor. Dev Biol *240*, 92–107.
Anbarci, D.N., McKey, J., Levic, D.S., Bagnat, M., and Capel, B. (2025). Rediscovering the rete ovarii, a secreting auxiliary structure to the ovary. Elife *13*.
Arango, N.A., Kobayashi, A., Wang, Y., Jamin, S.P., Lee, H.H., Orvis, G.D., and Behringer, R.R. (2008). A mesenchymal perspective of Müllerian duct differentiation and regression in Amhr2-lacZ mice. Mol Reprod Dev *75*, 1154–1162.
Bandiera, R., Vidal, V.P., Motamedi, F.J., Clarkson, M., Sahut-Barnola, I., von Gise, A., Pu, W.T., Hohenstein, P., Martinez, A., and Schedl, A. (2013). WT1 maintains adrenal-gonadal primordium identity and marks a population of AGP-like progenitors within the adrenal gland. Dev Cell *27*, 5–18.
Barbaux, S., Niaudet, P., Gubler, M.C., Grünfeld, J.P., Jaubert, F., Kuttenn, F., Fékété, C.N., Souleyreau-Therville, N., Thibaud, E., Fellous, M., *et al.* (1997). Donor splice-site mutations in WT1 are responsible for Frasier syndrome. Nat Genet *17*, 467–470.
Barrionuevo, F., Bagheri-Fam, S., Klattig, J.r., Kist, R., Taketo, M.M., Englert, C., and Scherer, G. (2006). Homozygous Inactivation of Sox9 Causes Complete XY Sex Reversal in Mice1. Biology of Reproduction *74*, 195–201.
Barsoum, I., and Yao, H.H. (2011). Redundant and differential roles of transcription factors Gli1 and Gli2 in the development of mouse fetal Leydig cells. Biol Reprod *84*, 894–899.
Barsoum, I.B., and Yao, H.H. (2010). Fetal Leydig cells: progenitor cell maintenance and differentiation. J Androl *31*, 11–15.
Bellutti, L., Abby, E., Tourpin, S., Messiaen, S., Moison, D., Trautmann, E., Guerquin, M.-J., Rouiller-Fabre, V., Habert, R., and Livera, G. (2019). Divergent Roles of CYP26B1 and Endogenous Retinoic Acid in Mouse Fetal Gonads. Biomolecules *9*, 536.
Berta, P., Hawkins, J.R., Sinclair, A.H., Taylor, A., Griffiths, B.L., Goodfellow, P.N., and Fellous, M. (1990). Genetic evidence equating SRY and the testis-determining factor. Nature *348*, 448–450.
Birk, O.S., Casiano, D.E., Wassif, C.A., Cogliati, T., Zhao, L., Zhao, Y., Grinberg, A., Huang, S., Kreidberg, J.A., Parker, K.L., *et al.* (2000). The LIM homeobox gene Lhx9 is essential for mouse gonad formation. Nature *403*, 909–913.
Boudreau, L., Robert, N.M., and Tremblay, J.J. (2026). Genetic ablation of the adult Leydig cell lineage reveals stem Leydig cells and a peritubular regenerative niche in mice. Endocrinology.
Boulanger, L., Pannetier, M., Gall, L., Allais-Bonnet, A., Elzaiat, M., Le Bourhis, D., Daniel, N., Richard, C., Cotinot, C., Ghyselinck, N.B., *et al.* (2014). FOXL2 is a female sex-determining gene in the goat. Curr Biol *24*, 404–408.
Bowles, J., Knight, D., Smith, C., Wilhelm, D., Richman, J., Mamiya, S., Yashiro, K., Chawengsaksophak, K., Wilson, M.J., Rossant, J., *et al.* (2006). Retinoid signaling determines germ cell fate in mice. Science *312*, 596–600.
Brennan, J., Tilmann, C., and Capel, B. (2003). Pdgfr-alpha mediates testis cord organization and fetal Leydig cell development in the XY gonad. Genes Dev *17*, 800–810.

Brown, P.R., Estermann, M.A., Gruzdev, A., Scott, G.J., Hagler, T.B., Ray, M.K., and Yao, H.H.-C. (2025). Generation of a Nr2f2-driven inducible Cre mouse to target interstitial cells in the reproductive system†. Biology of Reproduction *113*, 220–230.
Buehr, M., Gu, S., and McLaren, A. (1993). Mesonephric contribution to testis differentiation in the fetal mouse. Development *117*, 273–281.
Bullejos, M., and Koopman, P. (2001). Spatially dynamic expression of Sry in mouse genital ridges. Dev Dyn *221*, 201–205.
Bullejos, M., and Koopman, P. (2005). Delayed Sry and Sox9 expression in developing mouse gonads underlies B6-Y(DOM) sex reversal. Dev Biol *278*, 473–481.
Bunce, C., Barske, L., Zhang, G., and Capel, B. (2023). Biased precursor ingression underlies the center-to-pole pattern of male sex determination in mouse. Development *150*.
Byskov, A.G. (1974). Does the rete ovarii act as a trigger for the onset of meiosis? Nature *252*, 396–397.
Byskov, A.G. (1986). Differentiation of mammalian embryonic gonad. Physiol Rev *66*, 71–117.
Cai, H., Liu, B., Wang, H., Sun, G., Feng, L., Chen, Z., Zhou, J., Zhang, J., Zhang, T., He, M.*, et al.* (2020). SP1 governs primordial folliculogenesis by regulating pregranulosa cell development in mice. J Mol Cell Biol *12*, 230–244.
Callier, P., Calvel, P., Matevossian, A., Makrythanasis, P., Bernard, P., Kurosaka, H., Vannier, A., Thauvin-Robinet, C., Borel, C., Mazaud-Guittot, S.*, et al.* (2014). Loss of function mutation in the palmitoyl-transferase HHAT leads to syndromic 46,XY disorder of sex development by impeding Hedgehog protein palmitoylation and signaling. PLoS Genet *10*, e1004340.
Cameron, F.J., and Sinclair, A.H. (1997). Mutations in SRY and SOX9: testis-determining genes. Hum Mutat *9*, 388–395.
Capel, B. (2000). The battle of the sexes. Mech Dev *92*, 89–103.
Chaboissier, M.C., Kobayashi, A., Vidal, V.I., Lutzkendorf, S., Van De Kant, H.J., Wegner, M., De Rooij, D.G., Behringer, R.R., and Schedl, A. (2004). Functional analysis of Sox8 and Sox9 during sex determination in the mouse. Development *131*, 1891–1901.
Chassot, A.A., Ranc, F., Gregoire, E.P., Roepers-Gajadien, H.L., Taketo, M.M., Camerino, G., de Rooij, D.G., Schedl, A., and Chaboissier, M.C. (2008). Activation of beta-catenin signaling by Rspo1 controls differentiation of the mammalian ovary. Hum Mol Genet *17*, 1264–1277.
Chen, H., Wang, Y., Ge, R., and Zirkin, B.R. (2017). Leydig cell stem cells: Identification, proliferation and differentiation. Mol Cell Endocrinol *445*, 65–73.
Chen, P., Zirkin, B.R., and Chen, H. (2020). Stem Leydig Cells in the Adult Testis: Characterization, Regulation and Potential Applications. Endocr Rev *41*, 22–32.
Cheng, K., Seita, Y., Moriwaki, T., Noshiro, K., Sakata, Y., Hwang, Y.S., Torigoe, T., Saitou, M., Tsuchiya, H., Iwatani, C.*, et al.* (2022). The developmental origin and the specification of the adrenal cortex in humans and cynomolgus monkeys. Sci Adv *8*, eabn8485.
Clark, A.M., Garland, K.K., and Russell, L.D. (2000). Desert hedgehog (Dhh) gene is required in the mouse testis for formation of adult-type Leydig cells and normal development of peritubular cells and seminiferous tubules. Biol Reprod *63*, 1825–1838.

Combes, A.N., Lesieur, E., Harley, V.R., Sinclair, A.H., Little, M.H., Wilhelm, D., and Koopman, P. (2009a). Three-dimensional visualization of testis cord morphogenesis, a novel tubulogenic mechanism in development. Dev Dyn *238*, 1033–1041.
Combes, A.N., Wilhelm, D., Davidson, T., Dejana, E., Harley, V., Sinclair, A., and Koopman, P. (2009b). Endothelial cell migration directs testis cord formation. Dev Biol *326*, 112–120.
Cooke, C.B., and Moris, N. (2021). Tissue and cell interactions in mammalian PGC development. Development *148*.
Cool, J., DeFalco, T.J., and Capel, B. (2011). Vascular-mesenchymal cross-talk through Vegf and Pdgf drives organ patterning. Proc Natl Acad Sci U S A *108*, 167–172.
Coveney, D., Cool, J., Oliver, T., and Capel, B. (2008). Four-dimensional analysis of vascularization during primary development of an organ, the gonad. Proc Natl Acad Sci U S A *105*, 7212–7217.
Cowan, R.G., and Quirk, S.M. (2021). Cells responding to hedgehog signaling contribute to the theca of ovarian follicles. Reproduction *161*, 437–448.
Cui, S., Ross, A., Stallings, N., Parker, K.L., Capel, B., and Quaggin, S.E. (2004). Disrupted gonadogenesis and male-to-female sex reversal in Pod1 knockout mice. Development *131*, 4095–4105.
De Cian, M.C., Pauper, E., Bandiera, R., Vidal, V.P., Sacco, S., Gregoire, E.P., Chassot, A.A., Panzolini, C., Wilhelm, D., Pailhoux, E., *et al.* (2017). Amplification of R-spondin1 signaling induces granulosa cell fate defects and cancers in mouse adult ovary. Oncogene *36*, 208–218.
de Lau, W., Barker, N., Low, T.Y., Koo, B.K., Li, V.S., Teunissen, H., Kujala, P., Haegebarth, A., Peters, P.J., van de Wetering, M., *et al.* (2011). Lgr5 homologues associate with Wnt receptors and mediate R-spondin signalling. Nature *476*, 293–297.
de Lau, W., Peng, W.C., Gros, P., and Clevers, H. (2014). The R-spondin/Lgr5/Rnf43 module: regulator of Wnt signal strength. Genes Dev *28*, 305–316.
Defalco, T., Saraswathula, A., Briot, A., Iruela-Arispe, M.L., and Capel, B. (2013). Testosterone levels influence mouse fetal Leydig cell progenitors through notch signaling. Biol Reprod *88*, 91.
DeFalco, T., Takahashi, S., and Capel, B. (2011). Two distinct origins for Leydig cell progenitors in the fetal testis. Dev Biol *352*, 14–26.
Djari, C., Mayère, C., Guy, M., Perea-Gomez, A., Bellutti, L., Barreau, P., Ademi, H., Rozier, A., Martinez, A.S., Gibson, T.J., *et al.* (2026). Ovarian development is driven by early spatiotemporal priming of the coelomic epithelium. bioRxiv, 2025.2009.2024.678234.
Edson, M.A., Nagaraja, A.K., and Matzuk, M.M. (2009). The mammalian ovary from genesis to revelation. Endocr Rev *30*, 624–712.
Eicher, E.M., Washburn, L.L., Whitney, J.B., 3rd, and Morrow, K.E. (1982). Mus poschiavinus Y chromosome in the C57BL/6J murine genome causes sex reversal. Science *217*, 535–537.
Eliveld, J., van den Berg, E.A., Chikhovskaya, J.V., van Daalen, S.K.M., de Winter-Korver, C.M., van der Veen, F., Repping, S., Teerds, K., and van Pelt, A.M.M. (2019). Primary human testicular PDGFRalpha+ cells are multipotent and can be differentiated into cells with Leydig cell characteristics in vitro. Hum Reprod *34*, 1621–1631.

Escalante-Alcalde, D., and Merchant-Larios, H. (1992). Somatic and germ cell interactions during histogenetic aggregation of mouse fetal testes. Exp Cell Res *198*, 150–158.
Essafi, A., Webb, A., Berry, R.L., Slight, J., Burn, S.F., Spraggon, L., Velecela, V., Martinez-Estrada, O.M., Wiltshire, J.H., Roberts, S.G.*, et al.* (2011). A wt1-controlled chromatin switching mechanism underpins tissue-specific wnt4 activation and repression. Dev Cell *21*, 559–574.
Estermann, M.A., Grimm, S.A., Kitakule, A.S., Rodriguez, K.F., Brown, P.R., McClelland, K., Amato, C.M., and Yao, H.H. (2025). NR2F2 regulation of interstitial cell fate in the embryonic mouse testis and its impact on differences of sex development. Nat Commun *16*, 3987.
Feng, L., Wang, Y., Cai, H., Sun, G., Niu, W., Xin, Q., Tang, X., Zhang, J., Wang, C., Zhang, H.*, et al.* (2016). ADAM10-Notch signaling governs the recruitment of ovarian pregranulosa cells and controls folliculogenesis in mice. J Cell Sci *129*, 2202–2212.
Ferrari, S., Harley, V.R., Pontiggia, A., Goodfellow, P.N., Lovell-Badge, R., and Bianchi, M.E. (1992). SRY, like HMG1, recognizes sharp angles in DNA. Embo j *11*, 4497–4506.
Fortune, J.E., and Armstrong, D.T. (1977). Androgen production by theca and granulosa isolated from proestrous rat follicles. Endocrinology *100*, 1341–1347.
Franks, S., and Hardy, K. (2018). Androgen Action in the Ovary. Front Endocrinol (Lausanne) *9*, 452.
Fraser, E.A. (1950). The development of the vertebrate excretory system. Biol Rev Camb Philos Soc *25*, 159–187.
Fujimoto, Y., Tanaka, S.S., Yamaguchi, Y.L., Kobayashi, H., Kuroki, S., Tachibana, M., Shinomura, M., Kanai, Y., Morohashi, K., Kawakami, K.*, et al.* (2013). Homeoproteins Six1 and Six4 regulate male sex determination and mouse gonadal development. Dev Cell *26*, 416–430.
Garcia-Alonso, L., Lorenzi, V., Mazzeo, C.I., Alves-Lopes, J.P., Roberts, K., Sancho-Serra, C., Engelbert, J., Marečková, M., Gruhn, W.H., Botting, R.A.*, et al.* (2022). Single-cell roadmap of human gonadal development. Nature *607*, 540–547.
Ge, R.S., Dong, Q., Sottas, C.M., Papadopoulos, V., Zirkin, B.R., and Hardy, M.P. (2006). In search of rat stem Leydig cells: identification, isolation, and lineage-specific development. Proc Natl Acad Sci U S A *103*, 2719–2724.
Gonen, N., and Lovell-Badge, R. (2019). The regulation of Sox9 expression in the gonad. Curr Top Dev Biol *134*, 223–252.
Gregoire, E.P., De Cian, M.C., Migale, R., Perea-Gomez, A., Schaub, S., Bellido-Carreras, N., Stévant, I., Mayère, C., Neirijnck, Y., Loubat, A.*, et al.* (2023). The -KTS splice variant of WT1 is essential for ovarian determination in mice. Science *382*, 600–606.
Gregoire, E.P., Stevant, I., Chassot, A.A., Martin, L., Lachambre, S., Mondin, M., de Rooij, D.G., Nef, S., and Chaboissier, M.C. (2018). NRG1 signalling regulates the establishment of Sertoli cell stock in the mouse testis. Mol Cell Endocrinol *478*, 17–31.
Guahmich, N.L., Man, L., Wang, J., Arazi, L., Kallinos, E., Topper-Kroog, A., Grullon, G., Zhang, K., Stewart, J., Schatz-Siemers, N.*, et al.* (2023). Human theca arises from ovarian stroma and is comprised of three discrete subtypes. Commun Biol *6*, 7.
Gubbay, J., Collignon, J., Koopman, P., Capel, B., Economou, A., Munsterberg, A., Vivian, N., Goodfellow, P., and Lovell-Badge, R. (1990). A gene mapping to the sex-determining region of the mouse Y chromosome is a member of a novel family of embryonically expressed genes. Nature *346*, 245–250.
Gui, Y., and Yuan, S. (2021). Epigenetic regulations in mammalian spermatogenesis: RNA-m(6)A modification and beyond. Cell Mol Life Sci *78*, 4893–4905.
Hacker, A., Capel, B., Goodfellow, P., and Lovell-Badge, R. (1995). Expression of Sry, the mouse sex determining gene. Development *121*, 1603–1614.
Hammes, A., Guo, J.K., Lutsch, G., Leheste, J.R., Landrock, D., Ziegler, U., Gubler, M.C., and Schedl, A. (2001). Two splice variants of the Wilms' tumor 1 gene have distinct functions during sex determination and nephron formation. Cell *106*, 319–329.
Hancock, G.V., Wamaitha, S.E., Peretz, L., and Clark, A.T. (2021). Mammalian primordial germ cell specification. Development *148*.
Hanley, N.A., Ball, S.G., Clement-Jones, M., Hagan, D.M., Strachan, T., Lindsay, S., Robson, S., Ostrer, H., Parker, K.L., and Wilson, D.I. (1999). Expression of steroidogenic factor 1 and Wilms' tumour 1 during early human gonadal development and sex determination. Mech Dev *87*, 175–180.
Hanley, N.A., Hagan, D.M., Clement-Jones, M., Ball, S.G., Strachan, T., Salas-Cortés, L., McElreavey, K., Lindsay, S., Robson, S., Bullen, P., *et al.* (2000). SRY, SOX9, and DAX1 expression patterns during human sex determination and gonadal development. Mech Dev *91*, 403–407.
Hanley, N.A., Rainey, W.E., Wilson, D.I., Ball, S.G., and Parker, K.L. (2001). Expression profiles of SF-1, DAX1, and CYP17 in the human fetal adrenal gland: potential interactions in gene regulation. Mol Endocrinol *15*, 57–68.
Hao, H.X., Xie, Y., Zhang, Y., Charlat, O., Oster, E., Avello, M., Lei, H., Mickanin, C., Liu, D., Ruffner, H., *et al.* (2012). ZNRF3 promotes Wnt receptor turnover in an R-spondin-sensitive manner. Nature *485*, 195–200.
Harikae, K., Miura, K., Shinomura, M., Matoba, S., Hiramatsu, R., Tsunekawa, N., Kanai-Azuma, M., Kurohmaru, M., Morohashi, K., and Kanai, Y. (2013). Heterogeneity in sexual bipotentiality and plasticity of granulosa cells in developing mouse ovaries. J Cell Sci *126*, 2834–2844.
Harley, V.R., Clarkson, M.J., and Argentaro, A. (2003). The molecular action and regulation of the testis-determining factors, SRY (sex-determining region on the Y chromosome) and SOX9 [SRY-related high-mobility group (HMG) box 9]. Endocr Rev *24*, 466–487.
Harris, A., Siggers, P., Corrochano, S., Warr, N., Sagar, D., Grimes, D.T., Suzuki, M., Burdine, R.D., Cong, F., Koo, B.K., *et al.* (2018). ZNRF3 functions in mammalian sex determination by inhibiting canonical WNT signaling. Proc Natl Acad Sci U S A *115*, 5474–5479.
Hatano, O., Takakusu, A., Nomura, M., and Morohashi, K. (1996). Identical origin of adrenal cortex and gonad revealed by expression profiles of Ad4BP/SF-1. Genes Cells *1*, 663–671.
Hiramatsu, R., Harikae, K., Tsunekawa, N., Kurohmaru, M., Matsuo, I., and Kanai, Y. (2010). FGF signaling directs a center-to-pole expansion of tubulogenesis in mouse testis differentiation. Development *137*, 303–312.
Hiramatsu, R., Kanai, Y., Mizukami, T., Ishii, M., Matoba, S., Kanai-Azuma, M., Kurohmaru, M., Kawakami, H., and Hayashi, Y. (2003). Regionally distinct potencies of mouse XY genital ridge to initiate testis differentiation dependent on anteroposterior axis. Dev Dyn *228*, 247–253.

Hirshfield, A.N. (1991). Development of follicles in the mammalian ovary. Int Rev Cytol *124*, 43–101.
Hu, Y.C., Okumura, L.M., and Page, D.C. (2013). Gata4 is required for formation of the genital ridge in mice. PLoS Genet *9*, e1003629.
Ikeda, Y., Shen, W.H., Ingraham, H.A., and Parker, K.L. (1994). Developmental expression of mouse steroidogenic factor-1, an essential regulator of the steroid hydroxylases. Mol Endocrinol *8*, 654–662.
Imaimatsu, K., Uchida, A., Hiramatsu, R., and Kanai, Y. (2022). Gonadal Sex Differentiation and Ovarian Organogenesis along the Cortical-Medullary Axis in Mammals. Int J Mol Sci *23*.
Jameson, S.A., Lin, Y.T., and Capel, B. (2012a). Testis development requires the repression of Wnt4 by Fgf signaling. Dev Biol *370*, 24–32.
Jameson, S.A., Natarajan, A., Cool, J., Defalco, T., Maatouk, D.M., Mork, L., Munger, S.C., and Capel, B. (2012b). Temporal transcriptional profiling of somatic and germ cells reveals biased lineage priming of sexual fate in the fetal mouse gonad. PLoS Genet *8*, e1002575.
Jeanes, A., Wilhelm, D., Wilson, M.J., Bowles, J., McClive, P.J., Sinclair, A.H., and Koopman, P. (2005). Evaluation of candidate markers for the peritubular myoid cell lineage in the developing mouse testis. Reproduction *130*, 509–516.
Jeays-Ward, K., Dandonneau, M., and Swain, A. (2004). Wnt4 is required for proper male as well as female sexual development. Dev Biol *276*, 431–440.
Jeske, Y.W., Bowles, J., Greenfield, A., and Koopman, P. (1995). Expression of a linear Sry transcript in the mouse genital ridge. Nat Genet *10*, 480–482.
Karl, J., and Capel, B. (1998). Sertoli cells of the mouse testis originate from the coelomic epithelium. Developmental Biology *203*, 323–333.
Kato, T., Miyata, K., Sonobe, M., Yamashita, S., Tamano, M., Miura, K., Kanai, Y., Miyamoto, S., Sakuma, T., Yamamoto, T.*, et al.* (2013). Production of Sry knockout mouse using TALEN via oocyte injection. Scientific Reports *3*, 3136.
Katoh-Fukui, Y., Owaki, A., Toyama, Y., Kusaka, M., Shinohara, Y., Maekawa, M., Toshimori, K., and Morohashi, K. (2005). Mouse Polycomb M33 is required for splenic vascular and adrenal gland formation through regulating Ad4BP/SF1 expression. Blood *106*, 1612–1620.
Kay, R.G.G., Reeves, R., Siggers, P., Greenaway, S., Mallon, A.M., Wells, S., Koo, B.K., Mayère, C., Nef, S., Greenfield, A.*, et al.* (2024). Gonadal sex reversal at single-cell resolution in Znrf3-deficient mice. Development *151*.
Kilcoyne, K.R., Smith, L.B., Atanassova, N., Macpherson, S., McKinnell, C., van den Driesche, S., Jobling, M.S., Chambers, T.J., De Gendt, K., Verhoeven, G.*, et al.* (2014). Fetal programming of adult Leydig cell function by androgenic effects on stem/progenitor cells. Proc Natl Acad Sci U S A *111*, E1924–1932.
Kim, Y., Kobayashi, A., Sekido, R., DiNapoli, L., Brennan, J., Chaboissier, M.C., Poulat, F., Behringer, R.R., Lovell-Badge, R., and Capel, B. (2006). Fgf9 and Wnt4 act as antagonistic signals to regulate mammalian sex determination. PLoS Biol *4*, e187.
Klamt, B., Koziell, A., Poulat, F., Wieacker, P., Scambler, P., Berta, P., and Gessler, M. (1998). Frasier syndrome is caused by defective alternative splicing of WT1 leading to an altered ratio of WT1 +/-KTS splice isoforms. Hum Mol Genet *7*, 709–714.
Knarston, I.M., Pachernegg, S., Robevska, G., Ghobrial, I., Er, P.X., Georges, E., Takasato, M., Combes, A.N., Jørgensen, A., Little, M.H.*, et al.* (2020). An In Vitro

Differentiation Protocol for Human Embryonic Bipotential Gonad and Testis Cell Development. Stem Cell Reports *15*, 1377–1391.
Koizumi, M., Oyama, K., Yamakami, Y., Kida, T., Satoh, R., Kato, S., Hidema, S., Oe, T., Goto, T., Clevers, H., *et al.* (2015). Lgr4 controls specialization of female gonads in mice. Biol Reprod *93*, 90.
Koo, B.K., Spit, M., Jordens, I., Low, T.Y., Stange, D.E., van de Wetering, M., van Es, J.H., Mohammed, S., Heck, A.J., Maurice, M.M., *et al.* (2012). Tumour suppressor RNF43 is a stem-cell E3 ligase that induces endocytosis of Wnt receptors. Nature *488*, 665–669.
Koopman, P., Gubbay, J., Vivian, N., Goodfellow, P., and Lovell-Badge, R. (1991). Male development of chromosomally female mice transgenic for Sry. Nature *351*, 117–121.
Kreidberg, J.A., Sariola, H., Loring, J.M., Maeda, M., Pelletier, J., Housman, D., and Jaenisch, R. (1993). WT-1 is required for early kidney development. Cell *74*, 679–691.
Kulibin, A.Y., and Malolina, E.A. (2020). Formation of the rete testis during mouse embryonic development. Dev Dyn *249*, 1486–1499.
Kumar, D.L., and DeFalco, T. (2018). A perivascular niche for multipotent progenitors in the fetal testis. Nat Commun *9*, 4519.
Kumar, S., Chatzi, C., Brade, T., Cunningham, T.J., Zhao, X., and Duester, G. (2011). Sex-specific timing of meiotic initiation is regulated by Cyp26b1 independent of retinoic acid signalling. Nat Commun *2*, 151.
Kusaka, M., Katoh-Fukui, Y., Ogawa, H., Miyabayashi, K., Baba, T., Shima, Y., Sugiyama, N., Sugimoto, Y., Okuno, Y., Kodama, R., *et al.* (2010). Abnormal epithelial cell polarity and ectopic epidermal growth factor receptor (EGFR) expression induced in Emx2 KO embryonic gonads. Endocrinology *151*, 5893–5904.
Lardenois, A., Suglia, A., Moore, C.L., Evrard, B., Noël, L., Rivaud, P., Besson, A., Toupin, M., Blévinal, J., Dumortier, C., *et al.* (2026). Single-cell exploration of gonadal somatic cell lineage specification during human sex determination. Dev Cell *61*, 400–415.e406.
Lavery, R., Chassot, A.A., Pauper, E., Gregoire, E.P., Klopfenstein, M., de Rooij, D.G., Mark, M., Schedl, A., Ghyselinck, N.B., and Chaboissier, M.C. (2012). Testicular differentiation occurs in absence of R-spondin1 and Sox9 in mouse sex reversals. PLoS Genet *8*, e1003170.
Lavery, R., Lardenois, A., Ranc-Jianmotamedi, F., Pauper, E., Gregoire, E.P., Vigier, C., Moreilhon, C., Primig, M., and Chaboissier, M.C. (2011). XY Sox9 embryonic loss-of-function mouse mutants show complete sex reversal and produce partially fertile XY oocytes. Dev Biol *354*, 111–122.
Li, L., Lin, W., Wang, Z., Huang, R., Xia, H., Li, Z., Deng, J., Ye, T., Huang, Y., and Yang, Y. (2024). Hormone Regulation in Testicular Development and Function. International Journal of Molecular Sciences *25*, 5805.
Li, S.Y., Gu, X., Heinrich, A., Hurley, E.G., Capel, B., and DeFalco, T. (2021). Loss of Mafb and Maf distorts myeloid cell ratios and disrupts fetal mouse testis vascularization and organogenesis†. Biol Reprod *105*, 958–975.
Lin, Y.T., Barske, L., DeFalco, T., and Capel, B. (2017). Numb regulates somatic cell lineage commitment during early gonadogenesis in mice. Development *144*, 1607–1618.
Liu, C., Peng, J., Matzuk, M.M., and Yao, H.H. (2015). Lineage specification of ovarian theca cells requires multicellular interactions via oocyte and granulosa cells. Nat Commun *6*, 6934.

Liu, C., Rodriguez, K., and Yao, H.H. (2016). Mapping lineage progression of somatic progenitor cells in the mouse fetal testis. Development *143*, 3700–3710.
Liu, C.F., Bingham, N., Parker, K., and Yao, H.H. (2009). Sex-specific roles of beta-catenin in mouse gonadal development. Hum Mol Genet *18*, 405–417.
Lovell-Badge, R., Canning, C., and Sekido, R. (2002). Sex-determining genes in mice: building pathways. Novartis Found Symp *244*, 4–18; discussion 18–22, 35–42, 253–257.
Lovell-Badge, R., and Robertson, E. (1990). XY female mice resulting from a heritable mutation in the primary testis-determining gene, Tdy. Development *109*, 635–646.
Luo, X., Ikeda, Y., and Parker, K.L. (1994). A cell-specific nuclear receptor is essential for adrenal and gonadal development and sexual differentiation. Cell *77*, 481–490.
Maatouk, D.M., DiNapoli, L., Alvers, A., Parker, K.L., Taketo, M.M., and Capel, B. (2008). Stabilization of beta-catenin in XY gonads causes male-to-female sex-reversal. Hum Mol Genet *17*, 2949–2955.
Maatouk, D.M., Mork, L., Chassot, A.A., Chaboissier, M.C., and Capel, B. (2013). Disruption of mitotic arrest precedes precocious differentiation and transdifferentiation of pregranulosa cells in the perinatal Wnt4 mutant ovary. Dev Biol *383*, 295–306.
MacLean, G., Li, H., Metzger, D., Chambon, P., and Petkovich, M. (2007). Apoptotic extinction of germ cells in testes of Cyp26b1 knockout mice. Endocrinology *148*, 4560–4567.
Maekawa, M., Kamimura, K., and Nagano, T. (1996). Peritubular myoid cells in the testis: their structure and function. Arch Histol Cytol *59*, 1–13.
Magers, M.J., Udager, A.M., Chinnaiyan, A.M., French, D., Myers, J.L., Jentzen, J.M., McHugh, J.B., Heider, A., and Mehra, R. (2016). Comprehensive Immunophenotypic Characterization of Adult and Fetal Testes, the Excretory Duct System, and Testicular and Epididymal Appendages. Applied Immunohistochemistry & Molecular Morphology *24*, e50–e68.
Magoffin, D.A. (2005). Ovarian theca cell. Int J Biochem Cell Biol *37*, 1344–1349.
Magoffin, D.A., and Weitsman, S.R. (1994). Insulin-like growth factor-I regulation of luteinizing hormone (LH) receptor messenger ribonucleic acid expression and LH-stimulated signal transduction in rat ovarian theca-interstitial cells. Biol Reprod *51*, 766–775.
Maheshwari, A., and Fowler, P.A. (2008). Primordial follicular assembly in humans--revisited. Zygote *16*, 285–296.
Major, A.T., Estermann, M.A., and Smith, C.A. (2021). Anatomy, Endocrine Regulation and Embryonic Development of the Rete Testis. Endocrinology.
Malolina, E.A., and Kulibin, A.Y. (2019). The rete testis harbors Sertoli-like cells capable of expressing DMRT1. Reproduction *158*, 399–413.
Mandel, H., Shemer, R., Borochowitz, Z.U., Okopnik, M., Knopf, C., Indelman, M., Drugan, A., Tiosano, D., Gershoni-Baruch, R., Choder, M., *et al.* (2008). SERKAL syndrome: an autosomal-recessive disorder caused by a loss-of-function mutation in WNT4. Am J Hum Genet *82*, 39–47.
Mannan, M.A., and O'Shaughnessy, P.J. (1991). Steroidogenesis during postnatal development in the mouse ovary. J Endocrinol *130*, 101–106.
Mayere, C., Regard, V., Perea-Gomez, A., Bunce, C., Neirijnck, Y., Djari, C., Bellido-Carreras, N., Sararols, P., Reeves, R., Greenaway, S., *et al.* (2022). Origin, specification and differentiation of a rare supporting-like lineage in the developing mouse gonad. Sci Adv *8*, eabm0972.

Mayerhofer, A. (2013). Human testicular peritubular cells: more than meets the eye. Reproduction *145*, R107–116.
McCoshen, J.A. (1982). In vivo sex differentiation of congeneic germinal cell aplastic gonads. Am J Obstet Gynecol *142*, 83–88.
McCoshen, J.A. (1983). Quantitation of sex chromosomal influence(s) on the somatic growth of fetal gonads in vivo. Am J Obstet Gynecol *145*, 469–473.
McKey, J., Anbarci, D.N., Bunce, C., Ontiveros, A.E., Behringer, R.R., and Capel, B. (2022). Integration of mouse ovary morphogenesis with developmental dynamics of the oviduct, ovarian ligaments, and rete ovarii. Elife *11*.
McLaren, A. (2003). Primordial germ cells in the mouse. Dev Biol *262*, 1–15.
Merchant, H. (1975). Rat gonadal and ovarioan organogenesis with and without germ cells. An ultrastructural study. Dev Biol *44*, 1–21.
Migale, R., Neumann, M., Mitter, R., Rafiee, M.R., Wood, S., Olsen, J., and Lovell-Badge, R. (2024). FOXL2 interaction with different binding partners regulates the dynamics of ovarian development. Sci Adv *10*, eadl0788.
Migrenne, S., Pairault, C., Racine, C., Livera, G., Geloso, A., and Habert, R. (2001). Luteinizing hormone-dependent activity and luteinizing hormone-independent differentiation of rat fetal Leydig cells. Mol Cell Endocrinol *172*, 193–202.
Miyabayashi, K., Katoh-Fukui, Y., Ogawa, H., Baba, T., Shima, Y., Sugiyama, N., Kitamura, K., and Morohashi, K. (2013). Aristaless related homeobox gene, Arx, is implicated in mouse fetal Leydig cell differentiation possibly through expressing in the progenitor cells. PLoS ONE *8*, e68050.
Miyamoto, N., Yoshida, M., Kuratani, S., Matsuo, I., and Aizawa, S. (1997). Defects of urogenital development in mice lacking Emx2. Development *124*, 1653–1664.
Miyawaki, S., Kuroki, S., Maeda, R., Okashita, N., Koopman, P., and Tachibana, M. (2020). The mouse <i>Sry</i> locus harbors a cryptic exon that is essential for male sex determination. Science *370*, 121–124.
Molenaar, R., de Rooij, D.G., Rommerts, F.F., Reuvers, P.J., and van der Molen, H.J. (1985). Specific destruction of Leydig cells in mature rats after in vivo administration of ethane dimethyl sulfonate. Biol Reprod *33*, 1213–1222.
Molyneaux, K.A., Stallock, J., Schaible, K., and Wylie, C. (2001). Time-lapse analysis of living mouse germ cell migration. Dev Biol *240*, 488–498.
Moniot, B., Declosmenil, F., Barrionuevo, F., Scherer, G., Aritake, K., Malki, S., Marzi, L., Cohen-Solal, A., Georg, I., Klattig, J., *et al.* (2009). The PGD2 pathway, independently of FGF9, amplifies SOX9 activity in Sertoli cells during male sexual differentiation. Development *136*, 1813–1821.
Morais da Silva, S., Hacker, A., Harley, V., Goodfellow, P., Swain, A., and Lovell-Badge, R. (1996). Sox9 expression during gonadal development implies a conserved role for the gene in testis differentiation in mammals and birds. Nat Genet *14*, 62–68.
Mork, L., Maatouk, D.M., McMahon, J.A., Guo, J.J., Zhang, P., McMahon, A.P., and Capel, B. (2012). Temporal differences in granulosa cell specification in the ovary reflect distinct follicle fates in mice. Biol Reprod *86*, 37.
Motamedi, F.J., Badro, D.A., Clarkson, M., Lecca, M.R., Bradford, S.T., Buske, F.A., Saar, K., Hübner, N., Brändli, A.W., and Schedl, A. (2014). WT1 controls antagonistic FGF and BMP-pSMAD pathways in early renal progenitors. Nat Commun *5*, 4444.

Munger, S.C., Natarajan, A., Looger, L.L., Ohler, U., and Capel, B. (2013). Fine time course expression analysis identifies cascades of activation and repression and maps a putative regulator of Mammalian sex determination. PLoS Genet *9*, e1003630.
Nakata, H., and Iseki, S. (2019). Three-dimensional structure of efferent and epididymal ducts in mice. J Anat *235*, 271–280.
Neirijnck, Y., Sararols, P., Kühne, F., Mayère, C., Weerasinghe Arachchige, L.C., Regard, V., Nef, S., and Schedl, A. (2023). Single-cell transcriptomic profiling redefines the origin and specification of early adrenogonadal progenitors. Cell Rep *42*, 112191.
Nel-Themaat, L., Jang, C.W., Stewart, M.D., Akiyama, H., Viger, R.S., and Behringer, R.R. (2011). Sertoli cell behaviors in developing testis cords and postnatal seminiferous tubules of the mouse. Biol Reprod *84*, 342–350.
Nel-Themaat, L., Vadakkan, T.J., Wang, Y., Dickinson, M.E., Akiyama, H., and Behringer, R.R. (2009). Morphometric analysis of testis cord formation in Sox9-EGFP mice. Dev Dyn *238*, 1100–1110.
Ng, A., Tan, S., Singh, G., Rizk, P., Swathi, Y., Tan, T.Z., Huang, R.Y., Leushacke, M., and Barker, N. (2014). Lgr5 marks stem/progenitor cells in ovary and tubal epithelia. Nat Cell Biol *16*, 745–757.
Nicol, B., Estermann, M.A., Yao, H.H., and Mellouk, N. (2022). Becoming female: Ovarian differentiation from an evolutionary perspective. Front Cell Dev Biol *10*, 944776.
Nicol, B., and Yao, H.H. (2015). Gonadal Identity in the Absence of Pro-Testis Factor SOX9 and Pro-Ovary Factor Beta-Catenin in Mice. Biol Reprod *93*, 35.
Niu, W., and Spradling, A.C. (2020). Two distinct pathways of pregranulosa cell differentiation support follicle formation in the mouse ovary. Proc Natl Acad Sci U S A *117*, 20015–20026.
O'Shaughnessy, P.J., Baker, P.J., Monteiro, A., Cassie, S., Bhattacharya, S., and Fowler, P.A. (2007). Developmental changes in human fetal testicular cell numbers and messenger ribonucleic acid levels during the second trimester. J Clin Endocrinol Metab *92*, 4792–4801.
Omotehara, T., Wu, X., Kuramasu, M., and Itoh, M. (2020). Connection between seminiferous tubules and epididymal duct is originally induced before sex differentiation in a sex-independent manner. Dev Dyn *249*, 754–764.
Orth, J.M. (1982). Proliferation of Sertoli cells in fetal and postnatal rats: a quantitative autoradiographic study. Anat Rec *203*, 485–492.
Ostrer, H., Huang, H.Y., Masch, R.J., and Shapiro, E. (2008). A Cellular Study of Human Testis Development. Sexual Development *1*, 286–292.
Ottolenghi, C., Omari, S., Garcia-Ortiz, J.E., Uda, M., Crisponi, L., Forabosco, A., Pilia, G., and Schlessinger, D. (2005). Foxl2 is required for commitment to ovary differentiation. Hum Mol Genet *14*, 2053–2062.
Pailhoux, E., Vigier, B., Vaiman, D., Servel, N., Chaffaux, S., Cribiu, E.P., and Cotinot, C. (2002). Ontogenesis of female-to-male sex-reversal in XX polled goats. Dev Dyn *224*, 39–50.
Palmer, S.J., and Burgoyne, P.S. (1991). In situ analysis of fetal, prepuberal and adult XX----XY chimaeric mouse testes: Sertoli cells are predominantly, but not exclusively, XY. Development *112*, 265–268.
Pannetier, M., Tilly, G., Kocer, A., Hudrisier, M., Renault, L., Chesnais, N., Costa, J., Le Provost, F., Vaiman, D., Vilotte, J.L., *et al.* (2006). Goat SRY induces testis development in XX transgenic mice. FEBS Lett *580*, 3715–3720.

Parma, P., Radi, O., Vidal, V., Chaboissier, M.C., Dellambra, E., Valentini, S., Guerra, L., Schedl, A., and Camerino, G. (2006). R-spondin1 is essential in sex determination, skin differentiation and malignancy. Nat Genet *38*, 1304–1309.
Pelosi, E., Forabosco, A., and Schlessinger, D. (2015). Genetics of the ovarian reserve. Front Genet *6*, 308.
Pepling, M.E. (2012). Follicular assembly: mechanisms of action. Reproduction *143*, 139–149.
Perea-Gomez, A., Bellido Carreras, N., Dhellemmes, M., Tang, F., Le Gallo, C., and Chaboissier, M.C. (2025). NR2F2 is required in the embryonic testis for fetal Leydig cell development. Elife *14*.
Perrotta, G., Condrea, D., and Ghyselinck, N.B. (2025). Meiosis and retinoic acid in the mouse fetal gonads: An unforeseen twist. Curr Top Dev Biol *161*, 59–88.
Pierucci-Alves, F., Clark, A.M., and Russell, L.D. (2001). A developmental study of the Desert hedgehog-null mouse testis. Biol Reprod *65*, 1392–1402.
Pisarska, M.D., Barlow, G., and Kuo, F.T. (2011). Minireview: roles of the forkhead transcription factor FOXL2 in granulosa cell biology and pathology. Endocrinology *152*, 1199–1208.
Pitetti, J.L., Calvel, P., Romero, Y., Conne, B., Truong, V., Papaioannou, M.D., Schaad, O., Docquier, M., Herrera, P.L., Wilhelm, D., *et al.* (2013). Insulin and IGF1 receptors are essential for XX and XY gonadal differentiation and adrenal development in mice. PLoS Genet *9*, e1003160.
Pontiggia, A., Rimini, R., Harley, V.R., Goodfellow, P.N., Lovell-Badge, R., and Bianchi, M.E. (1994). Sex-reversing mutations affect the architecture of SRY-DNA complexes. Embo j *13*, 6115–6124.
Pontiggia, A., Whitfield, S., Goodfellow, P.N., Lovell-Badge, R., and Bianchi, M.E. (1995). Evolutionary conservation in the DNA-binding and -bending properties of HMG-boxes from SRY proteins of primates. Gene *154*, 277–280.
Qin, J., Tsai, M.J., and Tsai, S.Y. (2008). Essential roles of COUP-TFII in Leydig cell differentiation and male fertility. PLoS ONE *3*, e3285.
Qiu, C., Martin, B.K., Welsh, I.C., Daza, R.M., Le, T.M., Huang, X., Nichols, E.K., Taylor, M.L., Fulton, O., O'Day, D.R., *et al.* (2024). A single-cell time-lapse of mouse prenatal development from gastrula to birth. Nature *626*, 1084–1093.
Rahmoun, M., Lavery, R., Laurent-Chaballier, S., Bellora, N., Philip, G.K., Rossitto, M., Symon, A., Pailhoux, E., Cammas, F., Chung, J., *et al.* (2017). In mammalian foetal testes, SOX9 regulates expression of its target genes by binding to genomic regions with conserved signatures. Nucleic Acids Res *45*, 7191–7211.
Rastetter, R.H., Bernard, P., Palmer, J.S., Chassot, A.A., Chen, H., Western, P.S., Ramsay, R.G., Chaboissier, M.C., and Wilhelm, D. (2014). Marker genes identify three somatic cell types in the fetal mouse ovary. Dev Biol *394*, 242–252.
Richards, J.S., Ren, Y.A., Candelaria, N., Adams, J.E., and Rajkovic, A. (2017). Ovarian Follicular Theca Cell Recruitment, Differentiation, and Impact on Fertility: 2017 Update. Endocrine Reviews *39*, 1–20.
Richards, J.S., Ren, Y.A., Candelaria, N., Adams, J.E., and Rajkovic, A. (2018). Ovarian Follicular Theca Cell Recruitment, Differentiation, and Impact on Fertility: 2017 Update. Endocr Rev *39*, 1–20.
Richardson, N., Gillot, I., Gregoire, E.P., Youssef, S.A., de Rooij, D., de Bruin, A., De Cian, M.C., and Chaboissier, M.C. (2020). Sox8 and Sox9 act redundantly for ovarian-to-testicular fate reprogramming in the absence of R-spondin1 in mouse sex reversals. Elife *9*.
Roelen, B.A.J., and Chuva de Sousa Lopes, S.M. (2022). Stay on the road: from germ cell specification to gonadal colonization in mammals. Philos Trans R Soc Lond B Biol Sci *377*, 20210259.
Rossitto, M., Déjardin, S., Rands, C.M., Le Gras, S., Migale, R., Rafiee, M.R., Neirijnck, Y., Pruvost, A., Nguyen, A.L., Bossis, G., *et al.* (2022). TRIM28-dependent SUMOylation protects the adult ovary from activation of the testicular pathway. Nat Commun *13*, 4412.
Rotgers, E., Jorgensen, A., and Yao, H.H. (2018). At the Crossroads of Fate-Somatic Cell Lineage Specification in the Fetal Gonad. Endocr Rev *39*, 739–759.
Saitou, M., and Yamaji, M. (2012). Primordial germ cells in mice. Cold Spring Harb Perspect Biol *4*.
Sasaki, K., Oguchi, A., Cheng, K., Murakawa, Y., Okamoto, I., Ohta, H., Yabuta, Y., Iwatani, C., Tsuchiya, H., Yamamoto, T., *et al.* (2021). The embryonic ontogeny of the gonadal somatic cells in mice and monkeys. Cell Rep *35*, 109075.
Schmahl, J., and Capel, B. (2003). Cell proliferation is necessary for the determination of male fate in the gonad. Dev Biol *258*, 264–276.
Schmahl, J., Eicher, E.M., Washburn, L.L., and Capel, B. (2000). Sry induces cell proliferation in the mouse gonad. Development *127*, 65–73.
Schmahl, J., Rizzolo, K., and Soriano, P. (2008). The PDGF signaling pathway controls multiple steroid-producing lineages. Genes Dev *22*, 3255–3267.
Schmidt, D., Ovitt, C.E., Anlag, K., Fehsenfeld, S., Gredsted, L., Treier, A.C., and Treier, M. (2004). The murine winged-helix transcription factor Foxl2 is required for granulosa cell differentiation and ovary maintenance. Development *131*, 933–942.
Schnabel, C.A., Selleri, L., and Cleary, M.L. (2003). Pbx1 is essential for adrenal development and urogenital differentiation. Genesis *37*, 123–130.
Sekido, R., Bar, I., Narvaez, V., Penny, G., and Lovell-Badge, R. (2004). SOX9 is up-regulated by the transient expression of SRY specifically in Sertoli cell precursors. Dev Biol *274*, 271–279.
Sekido, R., and Lovell-Badge, R. (2008). Sex determination involves synergistic action of SRY and SF1 on a specific Sox9 enhancer. Nature *453*, 930–934.
Shen, Y.C., Shami, A.N., Moritz, L., Larose, H., Manske, G.L., Ma, Q., Zheng, X., Sukhwani, M., Czerwinski, M., Sultan, C., *et al.* (2021). TCF21(+) mesenchymal cells contribute to testis somatic cell development, homeostasis, and regeneration in mice. Nat Commun *12*, 3876.
Sinclair, A.H., Berta, P., Palmer, M.S., Hawkins, J.R., Griffiths, B.L., Smith, M.J., Foster, J.W., Frischauf, A.M., Lovell-Badge, R., and Goodfellow, P.N. (1990). A gene from the human sex-determining region encodes a protein with homology to a conserved DNA-binding motif. Nature *346*, 240–244.
Skinner, M.K., Tung, P.S., and Fritz, I.B. (1985). Cooperativity between Sertoli cells and testicular peritubular cells in the production and deposition of extracellular matrix components. J Cell Biol *100*, 1941–1947.
Smith, P., Wilhelm, D., and Rodgers, R.J. (2014). Development of mammalian ovary. J Endocrinol *221*, R145–161.
Souali-Crespo, S., Condrea, D., Vernet, N., Féret, B., Klopfenstein, M., Grandgirard, E., Alunni, V., Cerciat, M., Jung, M., Mayere, C., *et al.* (2023). Loss of NR5A1 in mouse

Sertoli cells after sex determination changes cellular identity and induces cell death by anoikis. Development *150*.
Stévant, I., Abberbock, E., Ridnik, M., Weiss, R., Swisa, L., Futtner, C.R., Maatouk, D.M., Lovell-Badge, R., Malysheva, V., and Gonen, N. (2025). The gene regulatory landscape driving mouse gonadal supporting cell differentiation. Sci Adv *11*, eadv1885.
Stevant, I., Kuhne, F., Greenfield, A., Chaboissier, M.C., Dermitzakis, E.T., and Nef, S. (2019). Dissecting Cell Lineage Specification and Sex Fate Determination in Gonadal Somatic Cells Using Single-Cell Transcriptomics. Cell Rep *26*, 3272–3283 e3273.
Stevant, I., and Nef, S. (2018). Single cell transcriptome sequencing: A new approach for the study of mammalian sex determination. Mol Cell Endocrinol *468*, 11–18.
Stevant, I., and Nef, S. (2019). Genetic Control of Gonadal Sex Determination and Development. Trends Genet *35*, 346–358.
Stevant, I., Neirijnck, Y., Borel, C., Escoffier, J., Smith, L.B., Antonarakis, S.E., Dermitzakis, E.T., and Nef, S. (2018). Deciphering Cell Lineage Specification during Male Sex Determination with Single-Cell RNA Sequencing. Cell Rep *22*, 1589–1599.
Sucheston, M.E., and Cannon, M.S. (1968). Development of zonular patterns in the human adrenal gland. J Morphol *126*, 477–491.
Suzuki, H., Kanai-Azuma, M., and Kanai, Y. (2015). From Sex Determination to Initial Folliculogenesis in Mammalian Ovaries: Morphogenetic Waves along the Anteroposterior and Dorsoventral Axes. Sex Dev *9*, 190–204.
Taelman, J., Czukiewska, S.M., Moustakas, I., Chang, Y.W., Hillenius, S., van der Helm, T., van der Meeren, L.E., Mei, H., Fan, X., and Chuva de Sousa Lopes, S.M. (2024). Characterization of the human fetal gonad and reproductive tract by single-cell transcriptomics. Dev Cell *59*, 529–544.e525.
Taketo, T., Saeed, J., Nishioka, Y., and Donahoe, P.K. (1991). Delay of testicular differentiation in the B6.YDOM ovotestis demonstrated by immunocytochemical staining for müllerian inhibiting substance. Dev Biol *146*, 386–395.
Tang, H., Brennan, J., Karl, J., Hamada, Y., Raetzman, L., and Capel, B. (2008). Notch signaling maintains Leydig progenitor cells in the mouse testis. Development *135*, 3745–3753.
Teerds, K.J., de Rooij, D.G., Rommerts, F.F., and Wensing, C.J. (1990). Development of a new Leydig cell population after the destruction of existing Leydig cells by ethane dimethane sulphonate in rats: an autoradiographic study. J Endocrinol *126*, 229–236.
Teerds, K.J., and Huhtaniemi, I.T. (2015). Morphological and functional maturation of Leydig cells: from rodent models to primates. Hum Reprod Update *21*, 310–328.
Tomaselli, S., Megiorni, F., De Bernardo, C., Felici, A., Marrocco, G., Maggiulli, G., Grammatico, B., Remotti, D., Saccucci, P., Valentini, F., *et al.* (2008). Syndromic true hermaphroditism due to an R-spondin1 (RSPO1) homozygous mutation. Hum Mutat *29*, 220–226.
Tomizuka, K., Horikoshi, K., Kitada, R., Sugawara, Y., Iba, Y., Kojima, A., Yoshitome, A., Yamawaki, K., Amagai, M., Inoue, A., *et al.* (2008). R-spondin1 plays an essential role in ovarian development through positively regulating Wnt-4 signaling. Hum Mol Genet *17*, 1278–1291.
Tung, P.S., Skinner, M.K., and Fritz, I.B. (1984). Cooperativity between Sertoli cells and peritubular myoid cells in the formation of the basal lamina in the seminiferous tubule. Ann N Y Acad Sci *438*, 435–446.

Uhlenhaut, N.H., Jakob, S., Anlag, K., Eisenberger, T., Sekido, R., Kress, J., Treier, A.C., Klugmann, C., Klasen, C., Holter, N.I., *et al.* (2009). Somatic sex reprogramming of adult ovaries to testes by FOXL2 ablation. Cell *139*, 1130–1142.
Vainio, S., Heikkila, M., Kispert, A., Chin, N., and McMahon, A.P. (1999). Female development in mammals is regulated by Wnt-4 signalling. Nature *397*, 405–409.
Val, P., Martinez-Barbera, J.P., and Swain, A. (2007). Adrenal development is initiated by Cited2 and Wt1 through modulation of Sf-1 dosage. Development *134*, 2349–2358.
Val, P., and Swain, A. (2010). Gene dosage effects and transcriptional regulation of early mammalian adrenal cortex development. Mol Cell Endocrinol *323*, 105–114.
Vidal, V.P., Chaboissier, M.C., de Rooij, D.G., and Schedl, A. (2001). Sox9 induces testis development in XX transgenic mice. Nat Genet *28*, 216–217.
Walters, K.A., and Handelsman, D.J. (2018). Role of androgens in the ovary. Mol Cell Endocrinol *465*, 36–47.
Wang, Q., Lan, Y., Cho, E.S., Maltby, K.M., and Jiang, R. (2005). Odd-skipped related 1 (Odd 1) is an essential regulator of heart and urogenital development. Dev Biol *288*, 582–594.
Wang, Y., Li, X., Li, H., and Ge, R.S. (2026). Current understanding of adult stem cells in Leydig cell biology. Curr Top Dev Biol *168*, 163–209.
Warr, N., Carre, G.A., Siggers, P., Faleato, J.V., Brixey, R., Pope, M., Bogani, D., Childers, M., Wells, S., Scudamore, C.L., *et al.* (2012). Gadd45gamma and Map3k4 interactions regulate mouse testis determination via p38 MAPK-mediated control of Sry expression. Dev Cell *23*, 1020–1031.
Wenzel, J.G., and Odend'hal, S. (1985). The mammalian rete ovarii: a literature review. Cornell Vet *75*, 411–425.
Wilhelm, D., and Englert, C. (2002). The Wilms tumor suppressor WT1 regulates early gonad development by activation of Sf1. Genes Dev *16*, 1839–1851.
Wilhelm, D., Hiramatsu, R., Mizusaki, H., Widjaja, L., Combes, A.N., Kanai, Y., and Koopman, P. (2007). SOX9 regulates prostaglandin D synthase gene transcription in vivo to ensure testis development. J Biol Chem *282*, 10553–10560.
Wilhelm, D., Martinson, F., Bradford, S., Wilson, M.J., Combes, A.N., Beverdam, A., Bowles, J., Mizusaki, H., and Koopman, P. (2005). Sertoli cell differentiation is induced both cell-autonomously and through prostaglandin signaling during mammalian sex determination. Dev Biol *287*, 111–124.
Wilhelm, D., Perea-Gomez, A., Newton, A., and Chaboissier, M.C. (2025). Gonadal sex determination in vertebrates: rethinking established mechanisms. Development *152*.
Wilhelm, D., Yang, J.X., and Thomas, P. (2013). Chapter Three - Mammalian Sex Determination and Gonad Development. In Current Topics in Developmental Biology, P. Thomas, ed. (Academic Press), pp. 89–121.
Yao, H.H., and Capel, B. (2002). Disruption of testis cords by cyclopamine or forskolin reveals independent cellular pathways in testis organogenesis. Dev Biol *246*, 356–365.
Yao, H.H., Whoriskey, W., and Capel, B. (2002). Desert Hedgehog/Patched 1 signaling specifies fetal Leydig cell fate in testis organogenesis. Genes Dev *16*, 1433–1440.
Yin, Q., and Spradling, A.C. (2025). Distinct waves of ovarian follicles contribute to mouse oocyte production. Elife *14*.
Zhao, L., and Koopman, P. (2012). SRY protein function in sex determination: thinking outside the box. Chromosome Res *20*, 153–162.

Zheng, W., Zhang, H., Gorre, N., Risal, S., Shen, Y., and Liu, K. (2014). Two classes of ovarian primordial follicles exhibit distinct developmental dynamics and physiological functions. Hum Mol Genet *23*, 920–928.
Zubair, M., Ishihara, S., Oka, S., Okumura, K., and Morohashi, K. (2006). Two-step regulation of Ad4BP/SF-1 gene transcription during fetal adrenal development: initiation by a Hox-Pbx1-Prep1 complex and maintenance via autoregulation by Ad4BP/SF-1. Mol Cell Biol *26*, 4111–4121.
Zubair, M., Parker, K.L., and Morohashi, K. (2008). Developmental links between the fetal and adult zones of the adrenal cortex revealed by lineage tracing. Mol Cell Biol *28*, 7030–7040.